\documentclass[gmd,manuscript]{copernicus}
\nolinenumbers

\begin{document}

\title{Evaluating Skill and Stability of ArchesWeather and ArchesWeatherGen under Multi-Decadal Climate Simulations}


\Author[1,2][singhrenu502@gmail.com]{Renu}{Singh}
\Author[3][robert.brunstein@gmail.com]{Robert}{Brunstein}
\Author[4]{Antonia}{Jost}
\Author[1]{Yana}{Hasson}
\Author[5]{Thomas}{Rackow}
\Author[2,6]{Claire}{Monteleoni}
\Author[5]{Christian}{Lessig}
\Author[1]{Guillaume}{Couiaron}


\affil[1]{Google DeepMind, Paris, France}
\affil[2]{INRIA, Paris, France}
\affil[3]{Otto von Guericke University Magdeburg, Magdeburg, Germany}
\affil[4]{University Potsdam, Potsdam, Germany}
\affil[5]{European Center for medium-range Weather Forecasts, Reading, United Kingdom}
\affil[6]{University of Colorado, Boulder, Colorado}

\equalcontrib{1,2}




\runningtitle{TEXT}

\runningauthor{TEXT}

\received{}
\pubdiscuss{} 
\revised{}
\accepted{}
\published{}


\firstpage{1}

\maketitle

\begin{abstract}
We evaluate the climate simulation capabilities of ArchesWeather and ArchesWeatherGen, two machine learning models originally trained for weather forecasting and evaluated up to a 10-day lead time. ArchesWeather is a deterministic model, while ArchesWeatherGen is a probabilistic flow-matching model leveraging ArchesWeather's forecasts, enabling ensemble based uncertainty quantification.
In this work, we adapt these models to act as forced atmospheric models by using additional conditioning on the monthly mean sea surface temperature (SST) and sea ice cover (SIC) as boundary conditions. 
In particular, we follow the AI Model Intercomparison Project (AIMIP) Phase 1 protocol, which, analogous to the Atmospheric Model Intercomparison Project (AMIP), proposes a standardized experimental setup to evaluate the climate skill of machine learning based forced atmospheric models.
We present a comprehensive evaluation of both models under these conditions, including comparison against numerical climate models, ablation studies that examine key design choices in the extension, and an analysis of forced versus unforced configurations. Despite being originally developed for weather forecasting, we demonstrate that forced configurations of ArchesWeather and ArchesWeatherGen produce stable long-term climate simulations, have a stable annual cycle, and capture the drift of many climate variables. The models faithfully reproduce ERA5's climatology, large-scale circulations and interannual variability, and they capture the tails of the distributions. 
\end{abstract}


\introduction  

Earth system models (ESMs) allow for the simulation of all aspects of the Earth system, and are used in particular to assess how the climate changes in response to anthropogenic activity (also known as \textit{forcings}). Climate simulations are an important tool for climate scientists to understand the climate response in a range of scenarios over the next 100 years, and enable downstream workflows for policy-making and infrastructure planning \citep{lee2023climate}. 

In ESMs, the dynamics of physical phenomena are modeled with partial differential equations (PDEs). Through the need for fine spatial discretization in solving PDEs, climate models require hours on thousands of supercomputer cores to simulate only 1 year of climate \citep{balaji2017cpmip, segura2025nextgems}. Thus, computation is a large bottleneck for climate projections, and it limits the spatiotemporal resolution these models can run as well as the number of ensemble members and scenario runs. For weather forecasting, data-driven models have enabled tremendous improvements in computational costs. They are also often more accurate than traditional numerical models such as IFS \citep{bi2022pangu, lam2023learning, price2025probabilistic, alet2025skillful}, due to training on some datasets informed by observations such as ERA5 \citep{hersbach2020era5}. 

Given the huge success in weather, data-driven methods have the potential to also greatly improve the computational efficiency of climate simulations. Efficient machine learning (ML) based climate emulation can enable better uncertainty estimation through large ensemble generation. ML models that learn the conditional climate response given anthropogenic activity have the potential to generate climate simulations in a much wider variety of scenarios than classical models, enabling counterfactual reasoning. They also have the potential to improve the accuracy of climate modeling, by better leveraging the very large amount of available Earth observation data.

Although climate models have initially been built on radiative balance, dynamical modeling has quickly taken a central place in long-term climate simulation. Hence, it is natural to try to adapt weather models that simulate the short-term dynamics of the atmosphere (a few hours to a few days) to long-term climate emulators by including the physics that drive the long-term climate evolution, such as radiative forcing, heat storage in the ocean, sea ice melt, and greenhouse gas (GHG) forcing. In modern physics based climate models, a dynamical model for the atmosphere is coupled with other models (for the ocean, sea ice, and biosphere). To study the atmospheric component in isolation, the AMIP project proposed using the historical sea surface temperature as boundary condition, enabling the investigation of the atmosphere's response to these forcings. Since data-driven models have emerged as a much cheaper alternative to NWP weather models, the AIMIP project was proposed to compare AI models, where the first stage is an equivalent evaluation setup to the AMIP project.

Extending ML weather models to forced atmospheric models is conceptually straightforward: the variables chosen as forcing can be used as additional input when training the weather model, turning it into a conditional weather model. In the case of AIMIP, the selected forcings are the monthly mean sea surface temperature (SST) and sea ice cover (SIC). The resulting conditioning model can then be used to study alternative scenarios, e.g. by considering a sea surface temperature that is 2\,\unit{K} or 4\,\unit{K} degrees warmer than in the historical period.

We chose to adapt the ArchesWeather and ArchesWeatherGen weather models, for their relatively cheap training cost compared to other AI models \citep{couairon2026archesweathergen}. ArchesWeather is a deterministic weather model trained at 1.5\unit{\degree} resolution with next-state prediction, conditioned on the two previous atmospheric states, with a lead time of 24\unit{h}. The model can then be used autoregressively to generate trajectories of future atmospheric states. An ensemble of four independently trained models improves accuracy, resulting in ArchesWeather-Mx4. ArchesWeatherGen is a flow matching based generative model trained to sample from the distribution of state residuals, i.e., the difference between true states and predictions from the deterministic ArchesWeather-Mx4. ArchesWeatherGen enables probabilistic weather forecasting.

We adapt ArchesWeather and ArchesWeatherGen into forced atmospheric models by retraining them in accordance with the AIMIP protocol. In particular, we changed the training period, moved to daily averaged atmospheric states, and added SST and SIC as conditioning variables. We then make 40 year rollouts of these models initialized from 01 October 1978 and investigate the long-term dynamics, climatology, distribution tails, and fidelity of the physical fields in these rollouts compared to the ERA5 reference.

\begin{figure}[ht]
    \centering
    \includegraphics[width=1.0\textwidth]{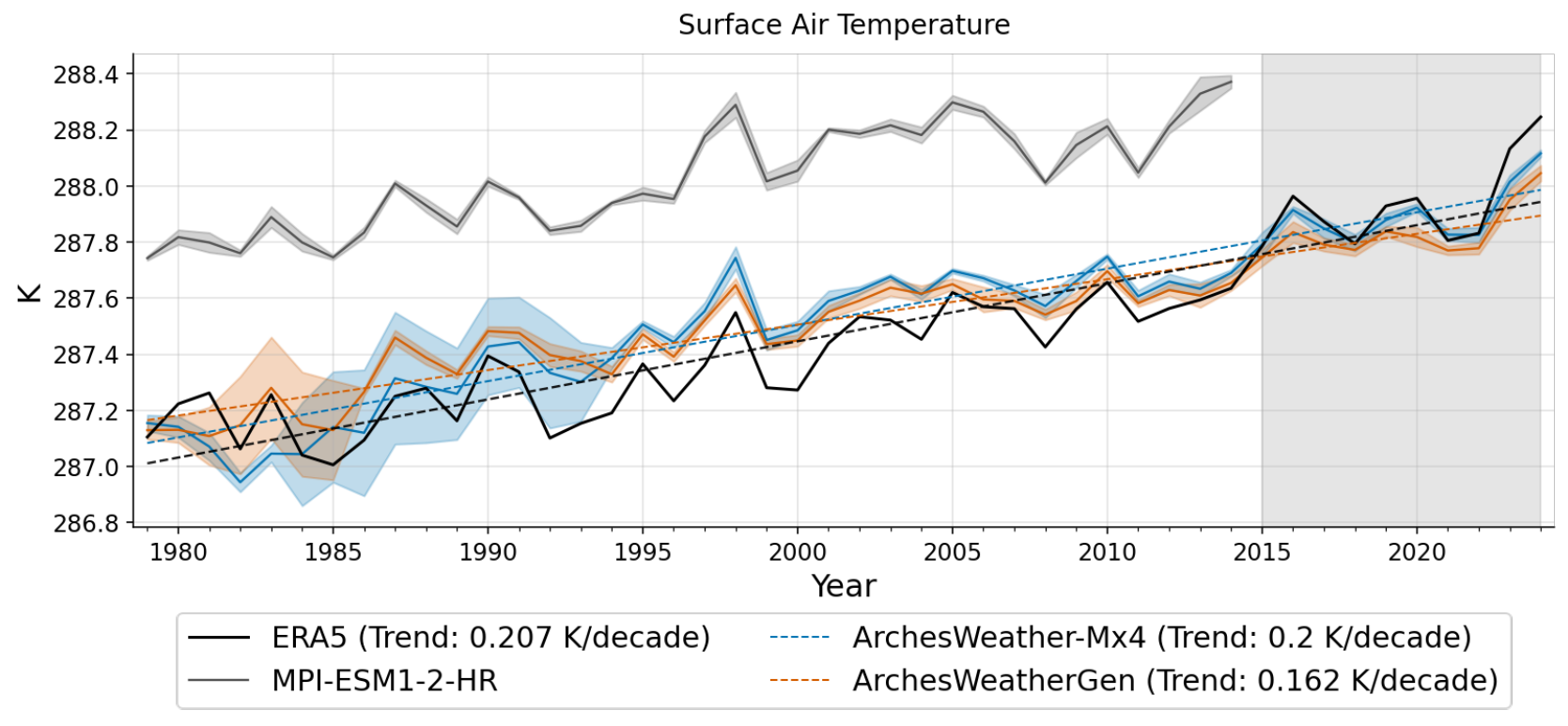}
    \caption{Global annual means of surface air temperature for ArchesWeatherGen, ArchesWeather, MPI-ESM1.2-HR and ERA5. The solid lines show ensemble means and the colored shaded areas show the standard deviation. The decadal trends of MPI-ESM1.2-HR and ArchesWeatherGen (dashed lines) over the full rollout period 1979--2024 are close to that of ERA5.}
    \label{fig:tas_aw_mpi_era} 
\end{figure}

We extensively evaluate our models on many different aspects of climate modeling. We first analyze the climatology of the simulated climate compared to the observed climate in ERA5 and demonstrate that our models closely capture the climatology (Fig. \ref{fig:tas_aw_mpi_era}). In particular, we analyze the seasonal cycle during the simulation period, the annual mean, and the spectral properties of the simulated data. Next, we evaluate the model's ability to reproduce some large-scale climate dynamics, such as the southern and northen annular modes, and the El-Niño Southern Oscillation.The South Asian monsoon has a significant impact on agriculture and civilian population in Asia, which we study with the Webster-Yang index. We measure the model's ability to simulate extreme events by comparing the return periods of climate events between our simulations and ERA5. We leave comparison with other machine learning models to the scope of the AIMIP collaboration, which is entirely focused around comparing ML based climate emulators.

Finally, we roll out models with a sea surface temperature 2\,\unit{K} or 4\,\unit{K} higher than in the historical period, and analyze the associated climate response. In particular, we analyze spatial bias maps between climate simulations subject to warmed forcings and ERA5 and analyze the zonal-mean profile of different variables. Although this setting is idealized, it is a first assessment of the suitability of our models to simulate climate scenarios. We find that our models are remarkably good given that the warmed forcings are out of distribution for the machine learning model and that forcings conditions are simply concatenated to the input. However, this response still differs from the expected climate response of such a warming. This is likely due to the lack of coupling between atmospheric models and ocean/sea ice variables.

\section{Related Work}
As stated previously, the data-driven paradigm led to the development of different models to generate climate simulations. Most of these new data-driven models are climate emulators. That is, they mimic the behavior of models that simulate the physics of the Earth system, in this case by learning the statistics from data. Models that simulate the climate are usually strongly tied to physical parameterizations which have to be tuned carefully. However, the distinctions between climate simulators and emulators are unclear, as data-driven models are more accurate without the need to build governing equations into them and only require a fraction of the compute cost.

An emulator model that showed early success is ACE2  \citep{watt2025ace2} which is based on the spherical fourier operator. ACE2 is trained either on data generated by the SHIELD model or on observational ERA5 data. An important contribution of the project was to provide stable simulations over 100 years at a fraction of the cost of physical climate models. A similar approach is presented in LUCIE \citep{LUCIE2025}, a lightweight emulator that can be trained on as few as 2 years of ERA5 data. The authors of LUCIE showed that their model is able to provide 100 year rollouts for 100 ensemble members, also with lower computational costs compared to common physics based models. 

The models mentioned so far are uncoupled models, i.e. they have no ocean component but are atmospheric emulators. An attempt to couple an atmospheric model with an ocean component is DLESYM \citep{Cresswell2025}, which consists of two U-Net models, one modeling the atmosphere at a lead time of 24 h and one modeling sea surface temperature with a lead time of 72 h. This allows the model, without external forcings, to capture the temporal structure of sea surface temperature, the only ocean variable used. 

Another approach called NeuralGCM (Neural General Circulation Model) is presented in \cite{kochkov2024neural}. NeuralGCM is a hybrid model that combines a general circulation model \citep{manabe1969climateGCM} with a ML based model. The authors showed stable decadal simulations and proved that combining ML based models with physical models yields more robust and stable simulations. Another hybrid approach was presented in \cite{clark2024ace2somcouplingmlatmospheric}, where the ACE model is extended by a slab-ocean model. This slab-ocean model receives prescribed ocean conditions but models the coupling between the atmosphere and sea surface temperature induced by radiative processes and greenhouse gases. 

A research direction distinct from the former ones is represented by \cite{brenowitz2025climatebottlegenerativefoundation} where the authors present a Climate in a Bottle (cBottle), that employs a two step process consisting of a non-autoregressive diffusion model that generates, conditioned on sea surface temperature and radiation, states of the atmosphere and downsamples the generated data to a finer resolution. Despite the fact that the model is not rolled out autoregressively, the model simulates temporally coherent patterns.

\section{Methods}
The goal of climate modeling is to model the distribution of future climate states (e.g. the value of atmospheric variables like wind and humidity, and ocean variables like sea surface temperature), under a given forcing scenario. Let $X_t$ denote the climate state at time $t$. We model climate state trajectories as sequences of evenly-spaced states $(X_{t+k\delta})_{1\le k \le K}$, where $\delta$ is the time step, which is fixed at $24h$ in the remainder of the paper.

\subsection{Data}\label{sec:data}

For training our models, we used daily averaged ERA5 data, regridded to 1\unit{\degree} resolution. Similar to standard practice in training weather models, we use 6 surface variables (surface air temperature, sea-level pressure, easwart wind at surface U and northward wind at surface V, sea surface temperature, sea ice concentration) and 6 upper air variables (air temperature, geopotential, specific humidity and the wind components U, V and W) on 13 pressure levels. To simulate coupling with the ocean, models are provided at training time with sea surface temperature and sea ice cover forcings, which are monthly means taken from ERA5 6-hourly data \citep{arcomano2025aimip}, and then regridded from the native grid (0.25\unit{\degree}) to 1\unit{\degree} using conservative regridding.

Rather than using instantaneous state variables as in ERA5, we pre-process the dataset by computing daily averaged physical variables, providing an easier target for the long-term dynamics by removing the daily cycle. Daily averages are generated from 6-hourly ERA5 using a rolling window of size 4 and stride 1. 

The dataset is split into 3 different periods: years 1979--2013 are used for training; 2014 is used for validation, and years 2015--2024 are reserved for testing the model. Notably, the data is non-stationary: there is a trend in increasing temperature, especially in the test period, reflecting global warming (see Fig. \ref{fig:tas_aw_mpi_era}).

\subsection{Training protocol}\label{section:pipeline}

We train the ArchesWeather and ArchesWeatherGen models using the same training protocol as in \cite{couairon2026archesweathergen} on the daily averaged data presented in section \ref{sec:data}. These models leverage a second-order Markovian approximation of the atmospheric dynamics, requiring only to learn the transition distribution $p(X_{t+\delta}|X_t, X_{t-\delta}, F_{t+\delta})$ conditioned on the forcings $F_{t+\delta}$. In more detail, the training protocol is as follows:

\begin{enumerate}
    \item A deterministic ArchesWeather model (85M parameters) is trained to predict the next state $X_{t+\delta}$ conditioned on the two previous states $X_t$ and $X_{t -\delta}$ as well as the conditioning variables $F_{t+\delta}$.
    \item the ArchesWeather-Mx4 model is the ensemble mean of four such models trained with different seeds for initialization.
    \item A dataset of normalized residuals is computed from the difference between the average predictions of ArchesWeather-Mx4 $f_\theta (X_t, F_{t+\delta})$ and the next state $X_{t+\delta}$ and then renormalized to unit variance: $r_{t+\delta} = (X_{t+\delta} - f_\theta (X_t,F_{t+\delta}))/\sigma$.
    \item A smaller flow-matching model ArchesWeatherGen (45M parameters) is trained on this dataset of residuals. This model can be seen as a probabilistic correction to ArchesWeather-Mx4's forecasts, that recovers a realistic atmospheric state from ArchesWeather-Mx4 smooth forecast and captures the stochasticity of the atmospheric dynamics.

\end{enumerate}

The deterministic ArchesWeather-M model was trained for 300K training steps, which took 1.5 days on 4 Google Cloud TPU v6 chips. The ArchesWeatherGen flow matching model was trained for 300K training steps and also took 1.5 days on 4 Google Cloud TPU v6 chips.

\textbf{Autoregressive rollouts.}\label{methods:rollouts} Given an initial state $(X_0, X_\delta)$ and a sequence of forcings $F_1, ... F_N$, the models are rolled out autoregressively to sample from the distribution

\begin{equation}\label{equation:autoregressive}
p(X_{2\delta:N\delta} | X_0, F_{2\delta:N\delta})
\approx \prod_{k=1..N} p(X_{k\delta+\delta}|X_{k\delta}, X_{k\delta-\delta}, F_{k\delta+\delta})
\end{equation}

Without forcings, the Markovian approximation would be valid at short lead times but not at climate time scales, because the long-term dynamics are not captured in the weather state $X_t$ and depend on additional information, e.g. the ocean variables or radiative forcings.
However, the forcings that we use as boundary condition do represent the slow moving ocean variables, which are required for modeling long-term climate with long-term variability. By capturing information relevant for long-term dynamics into conditioning variables, we can use the conditional Markovian decomposition, as the model will be informed about the short-term dynamics (through the current state $X_t$) as well as the long-term dynamics (through the forcings $F_{t+\delta}$).

We generate 45 years of climate rollouts by initializing the models with the observed daily average state from 01 October 1978 of ERA5 (which is not in the training dataset), and autoregressively predicting $X_{t+\delta}$ and feeding it back as input along with the forcing $F_{t+\delta}$, according to Equation \ref{equation:autoregressive}. For the forcings, we take either the historical monthly-mean values of SST and SIC, or the same values shifted uniformly on the entire globe by +2\,\unit{K} or +4\,\unit{K} in the warmed scenarios. A 45-year rollout ($\sim$16.5K inference steps per model) takes approximately 3.5 hours on a single TPU v6 chip for ArchesWeatherMx4 and 9 hours for ArchesWeatherGen. 

For rollouts of the ArchesWeather-Mx4 model, the average prediction of 4 single deterministic predictions is done for each autoregressive step. This differs from the original implementation, where, for weather rollouts up to 10 days, the different models were rolled out separately before averaging predictions at each step. For climate timescales, we found that in practice this makes a difference: averaging at each autoregressive step produces less biased rollouts. This could be explained by the fact that ensembling reduces the biases of the individual models, which otherwise can rapidly accumulate during long rollouts.

\textbf{Generating ensemble members.} To generate an ensemble of 5 members for the deterministic versions ArchesWeather and ArchesWeather-Mx4, we initialize the rollouts with slightly different initial conditions. We use initial conditions ($X_t$) from 29 September 1978, 30 September 1978, 1 October 1978, 2 October 1978, and 3 October 1978.

For rollouts of the stochastic model ArchesWeatherGen, we simply vary the input noise condition to generate rollouts from the same trained model (as in the original paper). The obvious advantage of the inherently stochastic model is that we can generate an arbitrary number of ensemble numbers with a fixed training budget. One difference with the original paper is that we do not scale input noise during model rollout, as we did not find improvement in diversity across all variables.

\subsection{Architecture}\label{section:architecture}
Both ArchesWeather and ArchesWeatherGen share the same transformer backbone, which improves upon the Pangu-Weather \citep{bi2022pangu} architecture with increased efficiency \citep{couairon2026archesweathergen}.
We make several modifications to the ArchesWeather backbone:
\begin{itemize}
  \item \textbf{Forcings}: The models take monthly mean forcings for sea surface temperature (SST) and sea ice cover (SIC) as additional input, by concatenating the forcings to the surface variables before embedding. 
  \item \textbf{Output SST/SIC}: We add daily SST and SIC as two extra output surface variables. This serves as sanity checks on how well the forcings are picked up by the model.
  \item \textbf{Adapted resolution}: Since the AIMIP protocol requires training at a resolution of 1\unit{\degree}, we change the patch size in the encoder and decoder for the 3D ViT backbone from 2x2x2 to 2x3x3, in order to maintain similar memory costs.
\end{itemize}
We also make several minor modifications, including NaN interpolation, masked loss, and time conditioning, which are described in more detail in \ref{appendix:architecture} along with an ablation study to justify our design choices.

\subsection{Evaluation}

In this section, we describe the different metrics that we use to evaluate our forced atmospheric simulations.

All models are evaluated by generating 5 trajectories, using either multiple initialization dates for the deterministic models, or sampling different noise for the stochastic model. All metrics are latitude-weighted using the latitude weights $w(i)$ defined in \cite{rasp2024weatherbench}, which account for the area distortion in the equirectangular maps. We compute metrics for each variable and pressure level separately.

\textbf{Climatology Ensemble Mean RMSE.} This metric measures whether we can capture the conditional expectation of future climate, given the SST and SIC forcings. We compute RMSE on an aggregated climatology rather than on daily snapshots because ERA5 represents only one possible climate trajectory conditioned on ocean surface forcings, and another climate trajectory would only match the states of ERA5 at weather timescales up to a few weeks, after which the anomalies of the trajectories become completely uncorrelated.

We define Climatology RMSE for a certain variable and pressure level as the following:
\begin{equation}
RMSE = \sqrt{\frac{1}{|G|} \sum_{i \in G} w(i) (\bar{f}_{i} - \bar{o}_{i})^2}
\end{equation} 
where $\bar{f}_{i}$ and $\bar{o}_{i}$ are the climatological means of the model forecasts and ground truth observations, respectively, for grid point $i \in G$ calculated as $\bar{x}_{i} = \frac{1}{|T|} \sum_{t \in T} \frac{1}{M} \sum_{m=1}^{M} x_{i,t,m}$ for all members of the ensemble $m \in \{1, \dots, M\}$ over a certain time period $t \in T$.

We compute a variant, monthly climatology RMSE, where we replace  $\bar{x}_{i}$ with $\bar{x}_{i,k} = \frac{1}{|T|} \sum_{t \in T} \frac{1}{M} \sum_{m=1}^{M} x_{i,k,t,m}$ for each month $k \in [\![1, 12]\!]$ and then average the squared difference over all $k$: 

Comparing monthly climatology ensures that we can capture seasonality. 
\begin{equation}
RMSE_{monthly} = \sqrt{\frac{1}{12} \sum_{k=1}^{12} \frac{1}{|G|} \sum_{i \in G} w(i) (\bar{f}_{i,k} - \bar{o}_{i,k})^2}
\end{equation}

\textbf{Cramér–von Mises distance.} To evaluate a model ensemble's ability to represent the full distribution of possible climate states, we contribute a way to compute a distance score. We compute cramér–von mises distance, energy distance between the predicted and reference distribution as the integral of the squared difference between the cumulative distributions of the forecasts $X$ and the ground truth $Y$:

\begin{equation}
CM(X,Y) = \int_{-\infty}^{\infty} (P[X \le y] - P[Y \le y])^2 \, dy
\end{equation} 

which, in practice, can be computed with i.i.d. predictions $X$, $X'$ and i.i.d. ground truth samples $Y$, $Y'$ as the following:
\begin{equation}
CM(X,Y) = E|X - Y| - \frac{1}{2} E|X - X'| - \frac{1}{2} E|Y - Y'|
\end{equation}

The continuous ranked probability score (CRPS), which is used to measure probabilistic weather forecasting skill \citep{rasp2024weatherbench} is a special case of this distance, when there is only one sample for ground truth $Y$. In that case, the third term cannot be estimated and is removed from CRPS. Thus, $CRPS(X,Y)$ is equal to $CM(X,Y)$ up to a scalar constant, which depends on the true distribution.

Since we only have one sample of the climate from ERA5, we consider each year of ERA5 as a proxy for an individual member prediction. This is based on the assumption that interannual variability provides sufficient proxy for the distribution of possible climate states, allowing us to compare our model ensemble against a proxy distribution.

For a given grid point $i \in G$ and day $d \in [[1, D]]$ (with $D \in (365, 366)$), we calculate the CM distance as the following, where $N_o$ is the number of years in the evalation period and $N_f = \textrm{num members} \cdot N_o$):
\begin{equation}
CM_{v, i,d} = \frac{1}{N_f N_o} \sum_{a=1}^{N_f} \sum_{b=1}^{N_o} |f_a - o_b| - \frac{1}{2 N_f(N_f-1)} \sum_{a=1}^{N_f} \sum_{a'=1}^{N_f} |f_a - f_{a'}| - \frac{1}{2 N_o(N_o-1)} \sum_{b=1}^{N_o} \sum_{b'=1}^{N_o} |o_b - o_{b'}|
\end{equation}

We then compute the CM distance for a certain variable $v$ by averaging over all grid points $i \in G$ and all days of the year $d \in [[1, D]]$:

\begin{equation}\label{equation:cramer}
CM_v(X, Y) = \frac{1}{D \cdot |G|} \sum_{d=1}^{D} \sum_{i \in G} w(i) \cdot CM_{v, i,d}
\end{equation}

\textbf{Qualitative evaluation} 

Besides comparing mean climatology and distributional skill, we further assess our models with respect to large scale climate patterns. 

As the most important mode of interannual variability, we analyze the El-Niño southern oscillation (ENSO) simulated by our model. We select the pressure based \textit{southern oscillation index} (SOI) as our models are forced by monthly mean sea surface temperature. The SOI is computed as the standardized difference of sea-level pressure anomalies $SOI = (A_{Tahiti} - A_{Darwin}) / \sigma_{month}$ between Tahiti (central pacific) and Darwin (western pacific) \citep{AQuantitativeEvaluationofENSOIndices}. Here, $A_x = (PSL_x - \overline{PSL}) / \sigma_{PSL} $ is the standardized anomaly at the respective location. 

To evaluate the models with respect to the annular modes, we employ empirical orthogonal function analysis (EOF) \citep{Hannachi_eof} . We therefore define the mean centered (anomalies) data of the variables of interest on the period 1981-2010. Following that, the unweighted anomalies of the target variable are regressed against the first loading pattern (the first eigenvector) of the EOFs and the computed slope is used to represent the regression pattern. To compare different models, we use a \textit{Taylor diagram} \citep{taylor_diag_2001}. The Taylor diagram is constructed by computing the Pearson-correlation coefficient (r-value), the centered RMSE and the standard deviation ratio between a ground truth pattern and the regression pattern computed from model simulations. For this, we select the best member per r-value as the member mean usually does not represent a distinct realization. This allows to compare different models reliably with respect to their skill reproducing such climate patterns.

\subsection{Reference Baselines}

\textbf{Numerical climate models.} We include a number of state-of-the-art physics based models, including MPI-ESM1.2-HR, CESM2-1-1-HR, GFDL-CM4-1x1, and HadGEM3-GC31-LL. We use AMIP runs which are most similar to our setup, where the atmospheric component is forced by SST and SIC rolled out from 1979 to 2014 \citep{eyring2016overview}. Since model output is available on the historical period until 2014, we can only compute train split metrics. The outputs are regridded with CDO to have 1\unit{\degree} resolution on both latitude and longitude, to match ERA5. Certain outputs is also available for the +4\,\unit{K} experiment called \textit{amip-p4k} \citep{webb2017cloud}, which we use as a reference for our out-of-distribution testing on warmed scenarios.

\textbf{Climatological Baseline.} For our quantitative metrics, we also compare to a simple climatological baseline which consists of the previous ten years of ERA5 before each evaluation split. For example, the climatological baseline for the test split (2015--2024) is the previous ten years (2005--2014).

\section{Results}

Here, we assess the ability of ArchesWeather and ArchesWeatherGen to act as forced atmospheric climate models and evaluate their simulations and discuss the models' climate response on warmed SST scenarios with a +2\,\unit{K} or +4\,\unit{K} bias uniformly added on the SST forcings.

\subsection{Climatology}

First, we assess how closely ArchesWeather and ArchesWeatherGen can replicate historical climatology by initializing the models from the atmospheric state on the first of October 1978 and running them forward to 2024 (the rollout method is detailed in Section \ref{section:pipeline}). We can compare climate predictions to historical ERA5 because (1) we remove the noise from interannual variability by averaging over space (for global trends) or time (for spatial bias) and (2) the models are constrained by historical ocean surface data, which also controls a large part of the interannual variability.

\textbf{Stability and global trend.} Figure ~\ref{fig:tas_aw_mpi_era} shows that the models' rollouts are stable and respond to ocean forcings over climate timescales, even when only trained on a prediction target with a 24-hour lead time. We compare the annual cycle of surface air temperature and observe that both ArchesWeather and ArchesWeatherGen produce a stable annual cycle with warming trends, similar to ERA5. 

However, we see that the trend increases significantly in the test period due to climate change, and the models fail to fully capture this increase in trend.

\begin{figure}[htbp]
    \centering
    \includegraphics[width=0.9\textwidth]{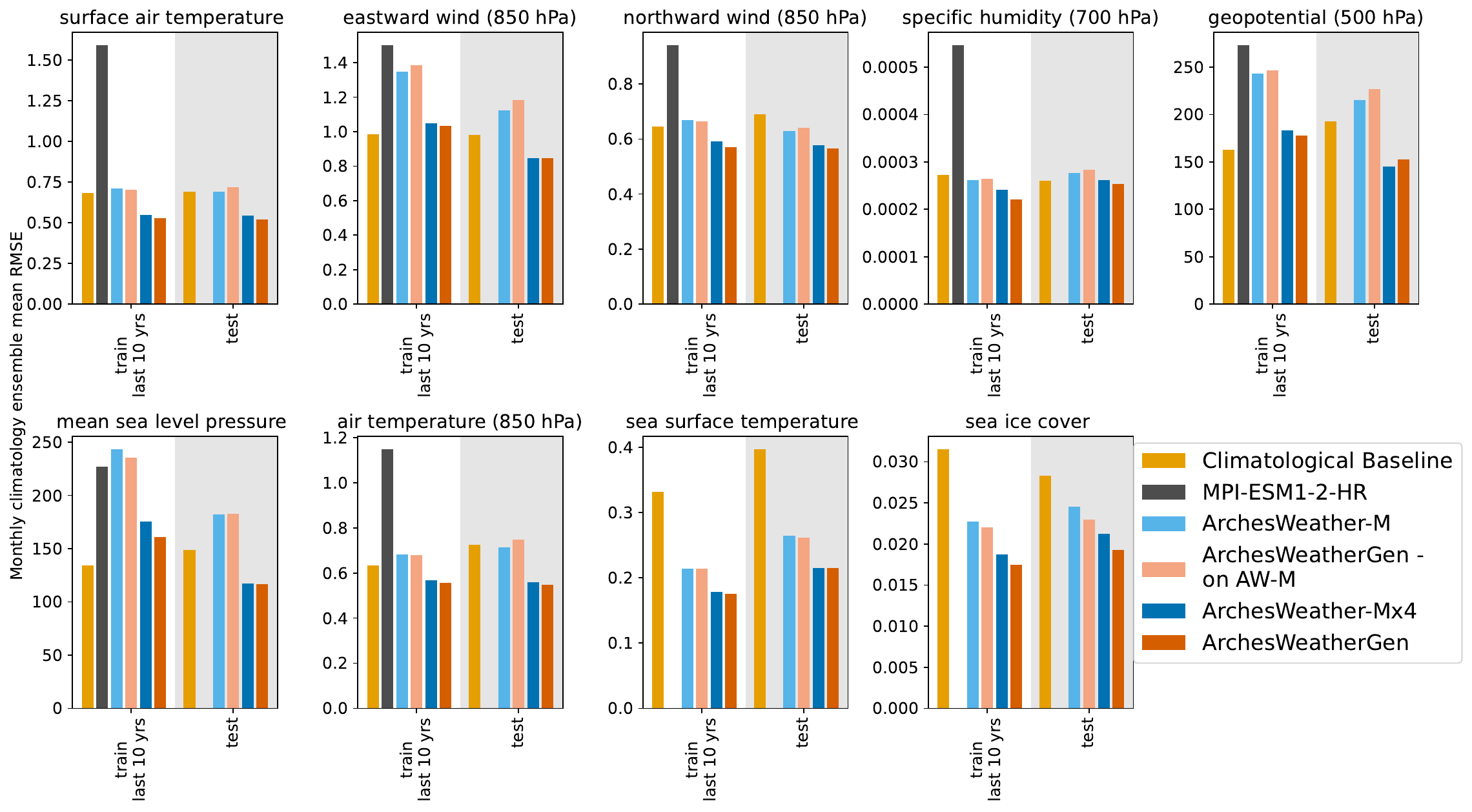}
    \caption{RMSE of monthly climatology for train (last 10 years only) and the test period. In general, the ensemble mean of ArchesWeatherMx4 and ArchesWeatherGen have similar scores, as expected. Furthermore, comparable test RMSE shows the models' ability to generalize to out of distribution climate.}
    \label{fig:rmse}
\end{figure}

\textbf{Spatial Bias.} In Fig.~\ref{fig:rmse}, we compare the monthly climatology RMSE of our ensembles. MPI-ESM1.2-HR has a larger bias compared to our models, indicating that the data-driven approach effectively utilizes the monthly mean-forcings. The similar RMSE for deterministic and generative versions is consistent with the fact that the generative model is expected to improve variability but not affect the mean state prediction. Furthermore, we compare RMSE on training and test periods to verify that the model is not overfitting to the training data and forcings. We use only the last ten years of the training period to fairly compare the error magnitude with that of the test period. 

Since ArchesWeather and ArchesWeatherGen are based on an ensemble mean of 4 independently trained models (see section \ref{section:pipeline} for details), we compare them against weaker versions of those models that only use a single deterministic model, referred to as \textit{ArchesWeatherM} and \textit{ArchesWeatherGen-on AW-M} respectively. 
\textit{ArchesWeather-M} contains a single neural network that can be more biased than an ensemble mean. This bias can translate into a bias for \textit{ArchesWeatherGen-on AW-M} that leverages \textit{ArchesWeatherM}'s predictions.
Sampling these two models remains the same. We observe that these weaker versions of the models have a larger bias than the full models that use an ensemble mean across four seeds. We conclude that using an average model across four seeds consistently reduces bias across variables, as was observed in \cite{couairon2026archesweathergen}.

\begin{figure}[htbp]
    \centering
    \includegraphics[width=1.0\textwidth]{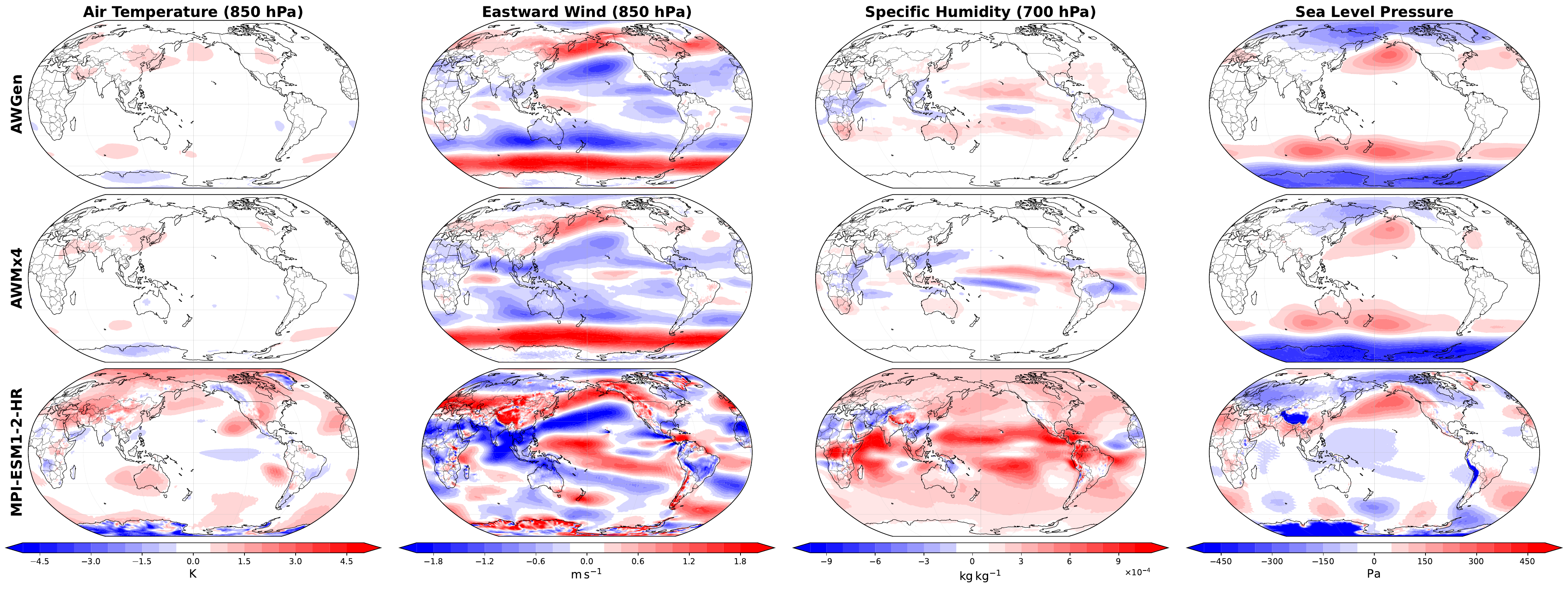}
    \caption{Bias maps of the mean climate for air temperature at 850\,\unit{hPa} (left), eastward wind at 850\,\unit{hPa} (middle left), specific humidity at 700\,\unit{hPa} (middle right) and sea sea-level pressure (right) averaged for all years from 1979 until 2014). The colorbars are scaled to 95th percentile values.
    Compared to our models, MPI-ESM1.2-HR shows substantially higher biases for air temperature at 850\,\unit{hPa} and specific humidity at 700\,\unit{hPa}. The different models show different biases, notably over the oceans in the tropics and mid-latitudes.}
    \label{fig:bias_maps} 
\end{figure}

In Fig.~\ref{fig:bias_maps}, we show spatial bias maps of the climatology for MPI-ESM1.2-HR, ArchesWeather, and ArchesWeatherGen averaged from 1979 to 2014 (due to the data available for MPI-ESM1.2-HR).
For air temperature at 850\,\unit{hPa} and specific humidity at 700\,\unit{hPa}, ArchesWeather (-0.005\,\unit{K} and 2.15e-5\,\unit{kg\,kg^{-1}}) and ArchesWeatherGen (-0.05\,\unit{K} and 4.17\,\unit{kg\,kg^{-1}}) exhibit significantly smaller biases than MPI-ESM1.2-HR (0.24\,\unit{K} and 2.63e-4\,\unit{kg\,kg^{-1}}) for all latitudes. 
For eastward winds at 850\,\unit{hPa}, our models ArchesWeather and ArchesWeatherGen exhibit larger biases near the eastward jet stream in the southern hemisphere in the 30° to 60° southern latitude band. For specific humidity at 700\,\unit{hPa}, simulations with ArchesWeather yield higher biases over the tropics, whereas for MPI-ESM1.2-HR the biases are distributed globally. 
In the case of sea-level pressure, ArchesWeather, and ArchesWeatherGen simulate biases in the higher latitude areas in the northern and southern hemispheres. MPI-ESM1.2-HR shows less bias over the latitudes close to the Antarctic but additional bias over the central latitudes. Notably, all three models show a similar bias over the northern Atlantic, influencing the northern annular mode.

\subsection{Distributional skill and  Spectral Energy Distribution}
Next, we evaluate the ability of ArchesWeather and ArchesWeatherGen to capture the climate distribution with 5 ensemble members, evaluated over the 10 years of the test period. We compare distributions separately for each physical variable and spatial location, considering the 5 predicted members for each year as the empirical distribution for that location, which is assumed to be stationary. This distribution with 50 data points is compared to the proxy reference distribution with 10 data points derived from the 10 years of ERA5 data during the test period. We use the Cramér–von Mises distance (CM, Equation \ref{equation:cramer}) as the distribution distance function. 

\begin{figure}[htpb!]
    \centering
    \includegraphics[width=0.9\textwidth]{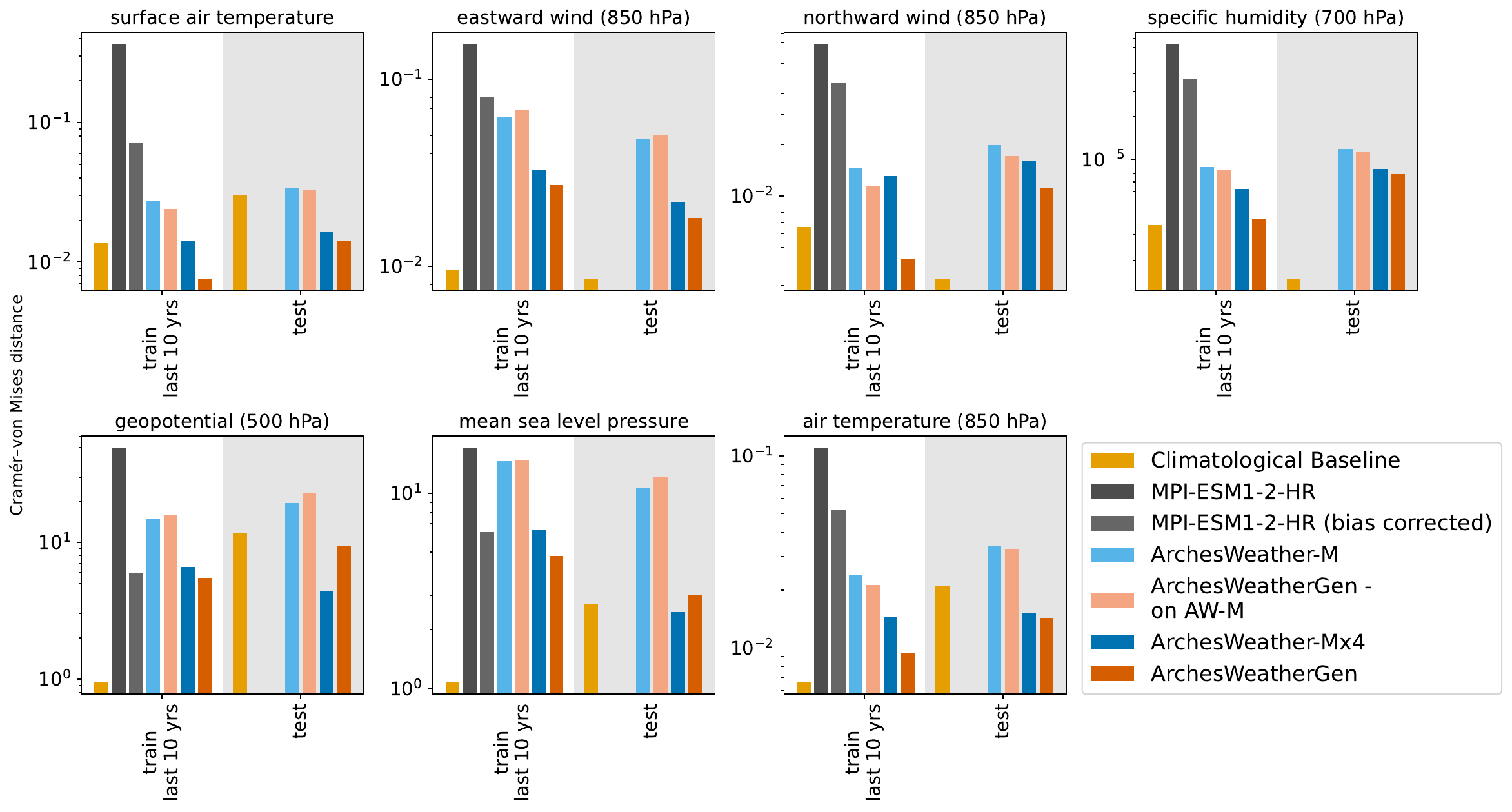}
    \caption{Cramér–von Mises distances show that overall, the final ArchesWeatherGen configuration, trained on the residuals w.r.t ArchesWeatherMx4, best reproduces the proxy distribution derived from the test period of ERA5, except for the smoother variables, geopotential at 500\,\unit{hPa} and sea-level pressure. We compare against the raw outputs and the mean corrected outputs of a physics based model MPI and a climatological baseline.}
    \label{fig:crps}
\end{figure}

We see in Fig.~\ref{fig:crps} that ArchesWeatherGen has the lowest CM distance against ERA5 amongst our model configurations, demonstrating the advantage of using generative modeling to approximate the distribution of climate states - in our case, daily averages of physical variables. Similarly to RMSE, the simulations based on an ensemble of 4 independently trained models achieve a better score compared to the models that use one single deterministic model (ArchesWeatherM and ArchesWeatherGen-on AW-M). Furthermore, we compare to a physics based model, MPI-ESM1.2-HR, which shows high CRPS, in part due to its strong biases. We, therefore, also compute CM distance using outputs that are corrected with the per-gridpoint mean from ERA5 over the evaluation period to show our models still hold a distributional advantage.

However, we see that ArchesWeatherGen has higher scores during the test period for two variables: geopotential and mean sea level pressure. We suspect that for these two variables, there is no high-frequency content, and ArchesWeatherGen adds too much variability to the deterministic prediction. We leave a detailed analysis for future work.

\begin{figure}[htbp!]
    \centering
    \includegraphics[width=1.0\textwidth]{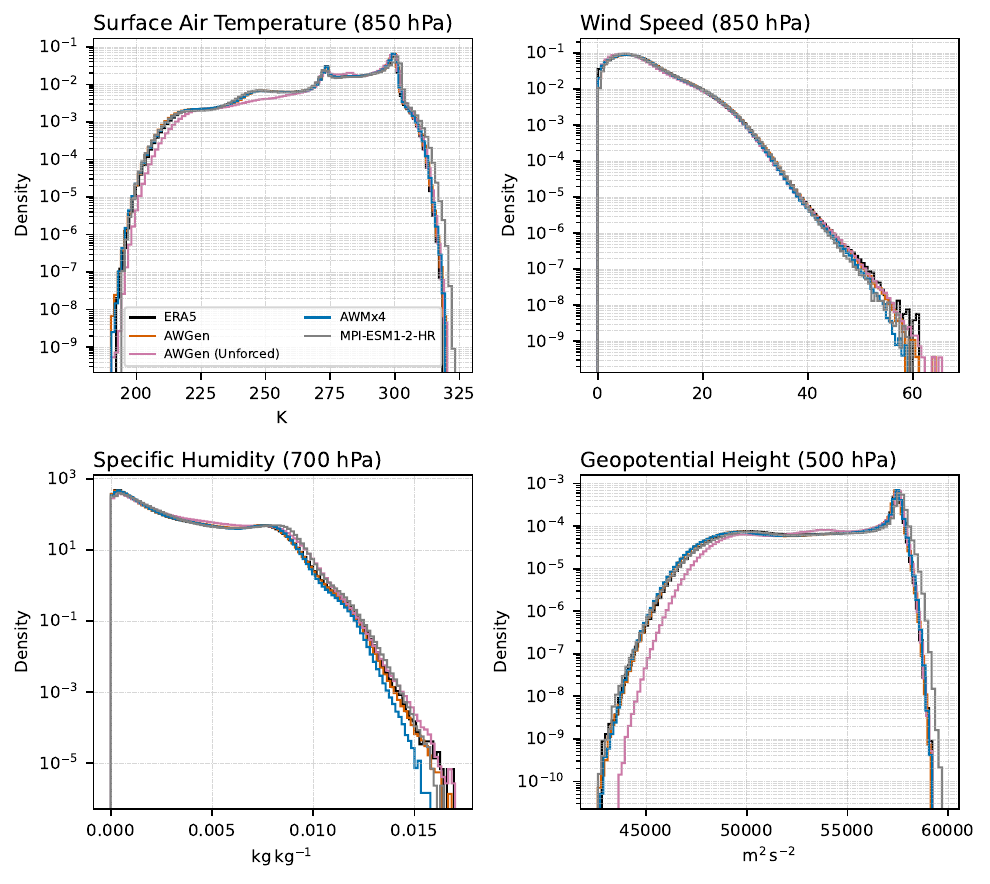}
    \caption{Histograms showcasing the distribution of air temperature at 850\,\unit{hPa}, wind speed at 850\,\unit{hPa}, specific humidity at 700\,\unit{hPa} and geopotential at 500\,\unit{hPa}. The advantage of the generative model is most pronounced for specific humidity but can also be observed for increasing magnitudes in wind speed at 850\,\unit{hPa}.}
    \label{fig:histograms}
\end{figure}

We also analyze the distributions of simulated field values qualitatively in figure \ref{fig:histograms}. The histograms are computed by pooling all members. Air temperature at 850\,\unit{hPa} shows that all models reproduce ERA5 distribution. However, the unforced model deviates around 250\,\unit{K}. For wind speed at 850\,\unit{hPa}, ArchesWeatherGen has a minor advantage compared to MPI-ESM1.2-HR and ArchesWeather. The advantage is more pronounced in the case of specific humidity, where ArchesWeather clearly misses the tail of the distribution, while ArchesWeatherGen is closest to the reference of ERA5. For the very smooth field of geopotential at 500\,\unit{hPa}, the choice of model makes no difference, although MPI-ESM1.2-HR overestimates higher potentials. 

\begin{figure}[htbp!]
    \centering
    \includegraphics[width=1.0\textwidth]{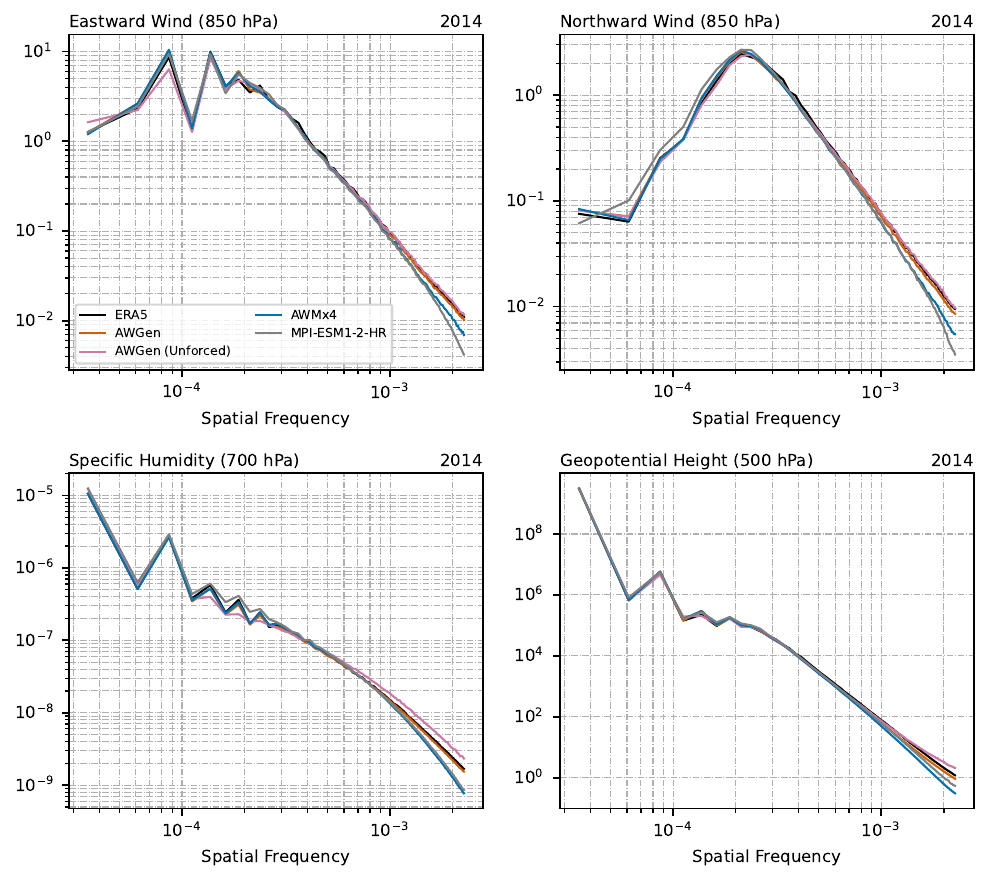}
    \caption{Spherical harmonics spectra of our models compared against ERA5 and MPI-ESM1.2-HR on four key variables. For all variables, ArchesWeatherGen shows advantages over both the numerical baseline model and ArchesWeather. The unforced configuration also maintains a spectrum similar to ERA55 }
    \label{fig:spherical_harmonics}
\end{figure}

Maintaining sharp predictions for long simulation periods is also indicative of the advantage of the generative model in capturing the full distribution rather than just the mean. In particular, for simulating extreme events, the correct representation of the physical spectrum at smaller scales is important. We visualize this with power spectra computed from the spherical harmonics transform in Fig. \ref{fig:spherical_harmonics}. ArchesWeather and MPI-ESM1.2-HR show relatively large variations compared to the ERA5 reference spectrum at shorter wavelengths. The discrepancy between ArchesWeather and ERA5 can be attributed to the mean-squared-error training objective that leads to smoothed predictions. In comparison, ArchesWeatherGen maintains a spectrum similar to ERA5 even at shorter wavelengths. This finding was demonstrated for weather forecasts in \cite{couairon2026archesweathergen}, and we demonstrate here that it is also valid for rollouts at climate timescales, which is an important feature of an ERA5-trained climate model. ArchesWeatherGen (unforced) maintains spectra similar to ArchesWeatherGen, but simulates higher spatial frequencies for wavelengths smaller than  500\,\unit{km} for specific humidity at 700\,\unit{hPa}. For the latter, it also produces a different spectrum in the range from 1000\,\unit{km} to 900\,\unit{km}, and for eastward wind at 850\,\unit{hPa}, it simulates a different spectrum for wavelengths 8000\,\unit{km} to 1000\,\unit{km}. Similar results also apply to northward wind at 850\,\unit{hPa}.

\subsection{Annular Modes}
\begin{figure}[htpb!]
    \centering
    \includegraphics[width=1.0\textwidth]{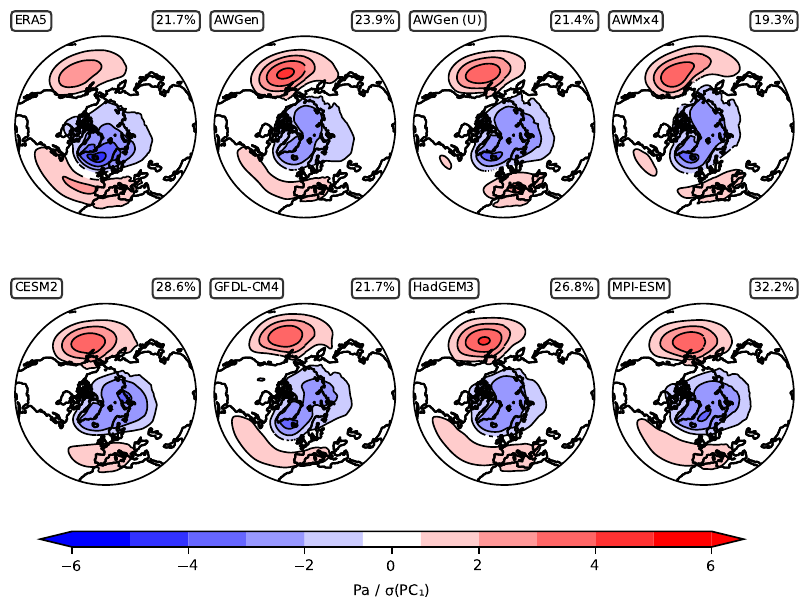}
    \caption{Leading empirical orthogonal functions of the northern hemisphere showing the tripole pattern with the low over Iceland and the highs over the Azores. The patterns is computed by projection sea-level pressure anomalies onto the leading mode EOF. We selected the best member for each model, measured by its r-value.}
    \label{fig:nam_pattern}
\end{figure}

Annular modes are large scale patterns of climate variability, following the latitude lines. The annular modes are the dominant patterns of variability in the northern and southern mid-to-high latitudes and are responsible for a large fraction of the total variance in geopotential and wind variable anomalies, thus significantly influencing the weather.

\begin{figure}[htpb!]
    \centering
    \includegraphics[width=1.0\textwidth]{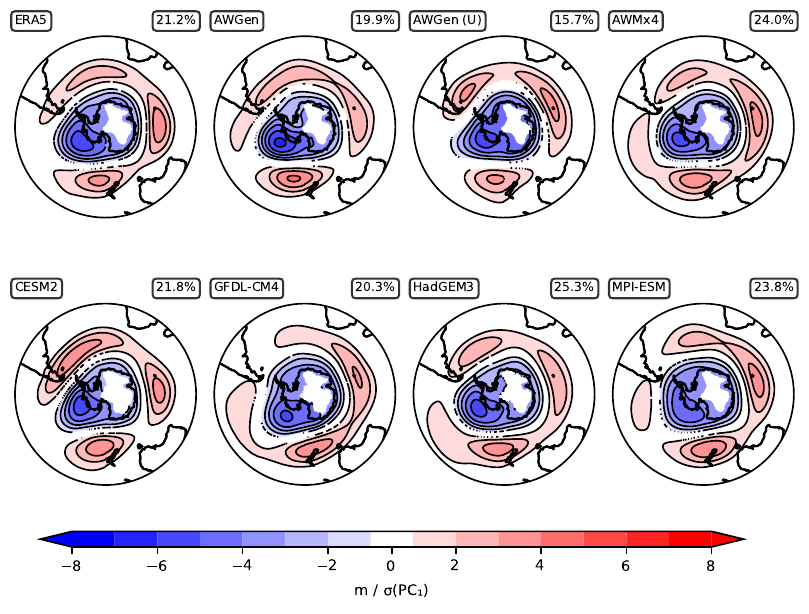}
    \caption{Leading empirical orthogonal functions of the southern hemisphere. All models show the zonally symmetric structure around the Antarctic. We applied orography masks to ERA5 and our models as they are present in the AMIP models.}
    \label{fig:sam_pattern}
   
\end{figure}

For the northern hemisphere, the Northern Annular Mode (NAM) \citep{thompson_1998_nam} describes the difference between the arctic and mid-latitude air pressure. To evaluate the reproduction of the NAM, we compute the empirical orthogonal functions of sea-level pressure anomalies from 20\unit{\degree\,N} to 90\unit{\degree\,N}. The centered data are latitude weighted, and the leading eigenvalue and corresponding eigenvector are computed using singular value decomposition. The unweighted anomalies are regressed against the EOF timeseries. We further compute r-values, centered RMSE, and standard deviation ratios against the ERA5 reference pattern for all models. The regression maps for the NAM are shown in Figure \ref{fig:nam_pattern}. ArchesWeatherGen shows a higher correlation in the northern Pacific and slightly underestimates the Icelandic low, but still closely matches the regression pattern of ERA5. ArchesWeather does not fully capture the correlation over the Atlantic ocean, similar to CESM2.  Qualitatively, the patterns reproduced by the CMIP6 models and our models are close to ERA5 \ref{fig:nam_pattern}. We also added the unforced configuration. There, the best member captures the anomalies well over the northern Atlantic and Pacific but shows a significantly lower regression slope over the Arctic, leading to the white space.  

The southern annular mode (SAM) \citep{thompson_2000_sam} dominates variability in the mid-to-high latitude southern hemisphere. The SAM causes a meridional (north-south) shift of the westerly wind belt. We compute the SAM similarly to the NAM but we use geopotential anomaly at 700\,\unit{hPa} instead of sea-level pressure. As three of the four numerical models apply orography masking, the patterns presented in Fig. \ref{fig:sam_pattern} have missing values in areas of higher elevation. We apply the same mask to ERA5 and our models for a fair comparison. The leading mode of ArchesWeatherGen explains a fraction of the total variance similar to ERA5 and shows a pattern that resembles the reference pattern. The pattern is slightly extended into the southern pacific and overestimates the magnitude of the higher elevated antarctic. Equal to the NAM, the unforced model is compared in addition to the forced configurations. We note that this configuration captures the pattern qualitatively relatively well.  

\begin{figure}[htpb!]
    \centering
    \includegraphics[width=1.0\textwidth]{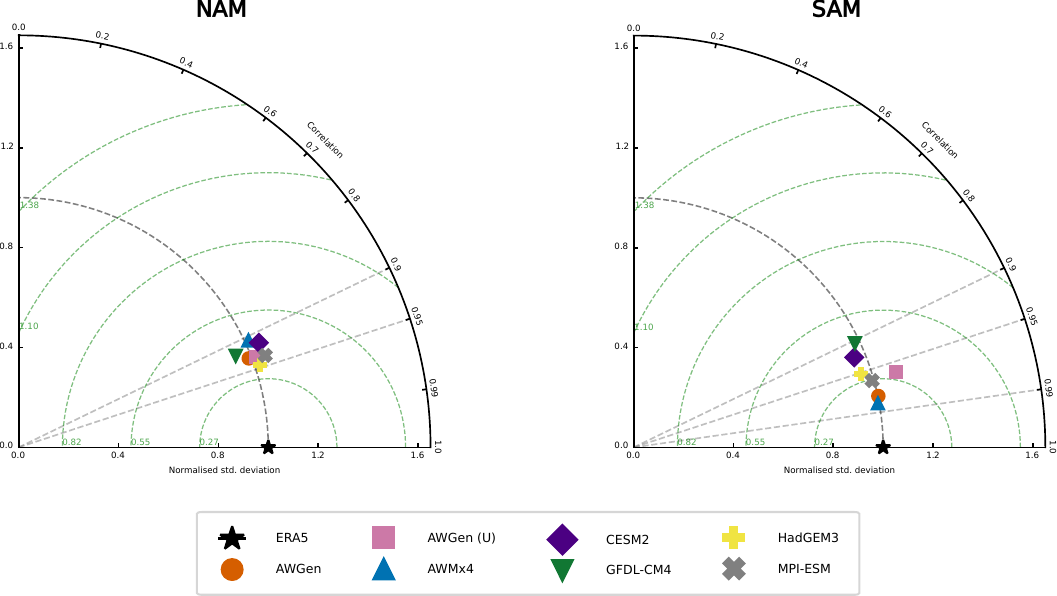}
    \caption{Taylor diagrams for the annular modes. For the SAM, ArchesWeatherGen and ArchesWeather are superior compared to the the numerical models. In case of the NAM, ArchesWeatherGen has a standard deviation ration close to 1 and ranks second best.  }
    \label{fig:taylor_annular_modes}
\end{figure}

We quantitatively assess the ability of our models in the Taylor diagrams shown in Fig. \ref{fig:taylor_annular_modes}. Again, we selected the best performing member for each model. All models show a strong correlation of more than 0.9 for the NAM and the SAM. ArchesWeather and ArchesWeatherGen show a standard deviation ratio close to $1.$ for the SAM and NAM. and thus faithfully represent the deviation from the mean. For the SAM, ArchesWeather and ArchesWeatherGen are the best performing models. We note that for the NAM, we found substantial deviations between the individual members of our models. While the best model has an r-value of $r \approx 0.93$, the model scoring worst has an r-value of $r \approx 0.59$, indicating a weaker correlation with ERA5. The mean value r of all members is $\bar{r} \approx 0.74$ and the standard deviation is $\bar{\sigma}_r \approx 0.11$. The reason for this might be related to the probabilistic nature of ArchesWeatherGen since for ArchesWeather we found that $\bar{r} \approx 0.81$ and $\bar{\sigma}_r \approx 0.09$. The physical models are more consistent compared to our models. The unforced configuration shows a strong performance for the NAM but is quantitatively weaker in case of the SAM, compared to the forced configurations.

\subsection{El Niño Southern Oscillation}\label{section:enso}
The El-Niño Southern Oscillation (ENSO) \citep{rasmusson_variations_1982} is an interannual mode of variability observed in the central and eastern pacific ocean. 
ENSO is a coupled climate process that involves changes in the atmosphere and the ocean surface. The ENSO is characterized by its three phases: El Niño (warmer than average SST) in the pacific ocean, La Niña (colder than average SST), and neutral. The appearance of El Niño or La Niña impacts many other regions of the world through teleconnection, and therefore, the faithful prediction of these events is a central task for climate models.

As our models are forced by monthly mean sea surface temperature, we evaluate the simulated ENSO pattern with the southern oscillation index (SOI) \citep{troup_soi_1965}. It expresses the atmospheric variability in the western and eastern tropical pacific in terms of sea-level pressure and thus better measures the ability of our model to take advantage of the effect of mean-forcings on other variables.

\begin{figure}[ht]
    \centering
    \includegraphics[width=1.0\textwidth]{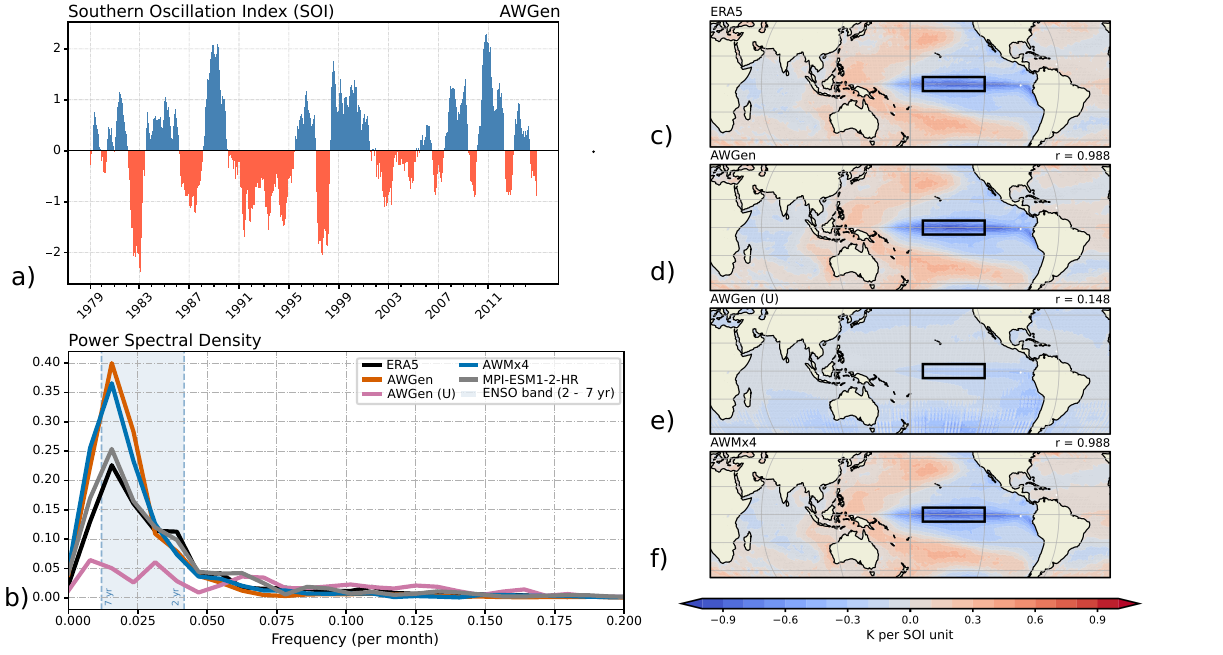}
    \caption{a) The computed Southern Oscillation Index for ArchesWeatherGen. In b), the power spectral density is computed for the unforced and the forced configuration of ArchesWeatherGen, the forced ArchesWeather model and ERA5. We observe that the forced model configurations have higher magnitudes in the range from 2 to 7 years (gray band) and underestimate the  higher frequencies slightly.  The unforced model is not able to reproduce the 2-7 year cycle of the ENSO. This is also reflected in the regression of the SOI against sea-surface-temperature, where ArchesWeatherGen (d) and ArchesWeather (f) have a r-value close to 1 and matches the pattern of ERA5 (c) closely, whereas the unforced configuration (e) is not able to reproduce it. In the Nino3.4 region (black box), the unforced model has a slope close to 0. The stippling pattern indicates statistical significance.}
    \label{fig:SOI}
\end{figure}

As the SOI is a station based index, we select the grid points closest to 
Tahiti, located at [149.6 \unit{\degree\,W}, 17.5\unit{\degree\,S}], and Darwin, located at [130.9 \unit{\degree\,E}, 12.4\unit{\degree\,S}]. 
In Fig. \ref{fig:SOI}a, we show the SOI computed from the ensemble mean of ArchesWeatherGen. 
Our generative model simulates a realistic ENSO cycle, and it reproduces strong El Niño events, e.g., from 1982--1983 and 1997--1998, as well as strong La Niña events, e.g., 1989 and 2011, matching the observed realizations of these patterns. The power spectral density of the SOI (\ref{fig:SOI}b) shows that most of the power for ArchesWeatherGen and ArchesWeather is located within the range of two to seven years. 
However, the magnitudes are larger than the ERA5 reference. We also added ArchesWeatherGen (unforced) as a baseline comparison. 
The unforced model has more power in the higher frequency range but  fails to capture slower oscillations. It is unable to capture the slower evolving dynamics of sea surface temperature that strongly influence the ENSO cycle. A reason for this might be related to the forecast lead time, as sea surface temperature is predicted at a 24 hour lead time, which makes it hard to capture the slower evolving field. 
We further regressed the SOI against sea surface temperature. 
The forced versions of ArchesWeatherGen, figure \ref{fig:SOI} d), and ArchesWeather ,figure \ref{fig:SOI} f), obtain an r-value close to $1.$ and reproduce the antipodal pattern similar to the ERA5 reference, depicted in figure \ref{fig:SOI} c).
Given its lack of skill in predicting long-term correlations, we can observe that the unforced model has a low r-value and is not able to reproduce the pattern expected around the Niño 3.4 region (indicated by the black box) as can be observed in figure \ref{fig:SOI} d).

\subsection{Indian Monsoon}
In this section, we inspect the ability of our models to reproduce the south Asian monsoon, which is responsible for up to 80\% of all precipitation in India and therefore provides most of the drinking water for the south Asian region and significantly influences agriculture and the economy \citep{turner2012climateMonsoon}.

\begin{figure}[htpb!]
    \centering
    \includegraphics[width=1.0\textwidth]{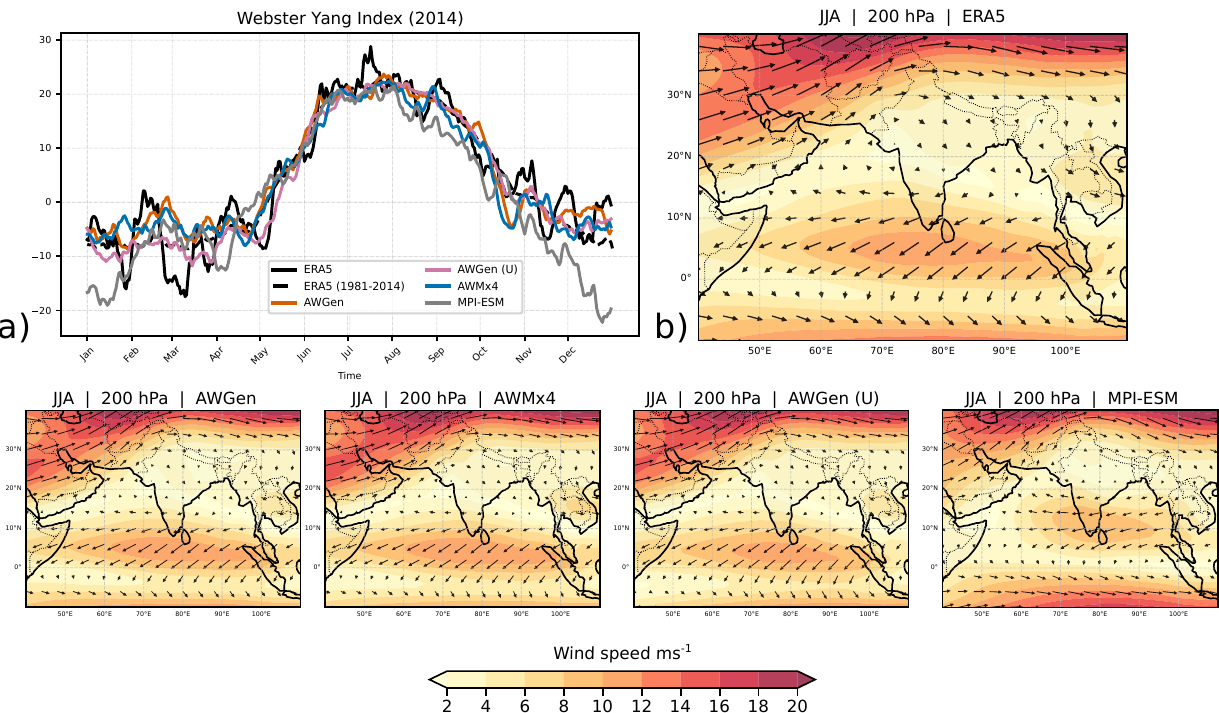}
    \caption{The Webster-Yang monsoon index a) computed for ArchesWeather, ArchesWeatherGen and MPI-ESM1.2-HR and plotted against the ERA5 reference. We also plot the average of ERA5 over a baseline period of 30 years. We observe great agreement between our models and the ERA5 reference. This is further strengthened by the wind vector fields at 200\,\unit{hPa}, where our models show a similar pattern to the ERA5 reference whereas MPI-ESM1.2-HR shows a northward shift of the maximum wind velocity over the Indian ocean and the southern part of India. The unforced configuration shows similar results to the forced ones ut with weaker magnitudes close to the African coast}
    \label{fig:wyi}
\end{figure}

As our models do not predict precipitation or outgoing long-wave radiation, we computed the Webster-Yang monsoon index (WYI) \citep{WebsterEnsoMonsoon}, which is a circulation index defined as the temporal mean zonal wind shear between zonal winds at 850\,\unit{hPa} and 250\,\unit{hPa} over south Asia (40\degree\,\unit{E}--110\degree\,\unit{E}, 0\degree--20\degree\,\unit{N}). Higher values indicate a stronger monsoon, while lower values indicate a weaker monsoon. 
Although not always highly correlated with precipitation based indices \citep{wang1999choiceMonssonIndex}, the WYI also provides insight into higher pressure levels, complementing our other evaluations. In Fig. \ref{fig:wyi}, the computed indices are shown for ArchesWeather, ArchesWeatherGen, MPI-ESM1.2-HR, and ERA5, as well as the unforced configuration of ArchesWeatherGen. We selected 2014 as our target year and plotted a 30-year mean for ERA5 as a reference. We observe that all models closely match the index of ERA5 during the south Asian summer monsoon period. However, during the winter months, MPI-ESM1.2-HR simulates a weaker index compared to ERA5, while the forced and unforced configurations remain close to the reference index.

We also compare the wind vector fields of the three models with the ERA5 reference in \ref{fig:wyi}. ArchesWeatherGen, ArchesWeather, and MPI-ESM1.2-HR show strong westerly winds over the Tibetan plateau and easterly winds of high magnitude over the Pacific Ocean with a clear ridge-like structure that leads to a strong circulation system over southern Asia and India in particular. However, while the monsoon indices are similar for the south Asian monsoon period, the vector fields also show distinct patterns when comparing ArchesWeather and ArchesWeatherGen against MPI-ESM1.2-HR. Although our models are close to the ERA5 reference, MPI-ESM1.2-HR shows a northward shift of the wind vector field. This also indicates that our models are able to simulate lower pressure-level variables in good agreement with the reference climatology, although they are less weighted in the loss computation during training. Strikingly, also the unforced model is able to model the monsoon winds reliably even after 36 years of rollout. The ridge-like structure over the pacific ocean is similar to the one of ERA5 with minor differences close to the African coast.

\subsection{Extremes}
The prediction of extreme weather events is an important aspect of climate simulations. To evaluate the ability of our models to represent such extremes, we computed the return periods for wind velocity at 850\,\unit{hPa} and surface air temperature. Return periods are computed using the Weibull plotting formula and 95\,\unit{\%} confidence intervals are derived from fitting a distribution function to the extremes. We obtain return periods longer than 45 years (1979--2024) by pooling model members using a stationarity assumption. For wind speed, computed as $v = \sqrt{u_x^2 + u_y^2}$, this assumption is valid. For surface air temperature, this assumption is violated due to the global warming trend. Nevertheless, pooling allows us to estimate return periods more accurately.

\begin{figure}[htpb!]
    \centering
    \includegraphics[width=1.0\textwidth]{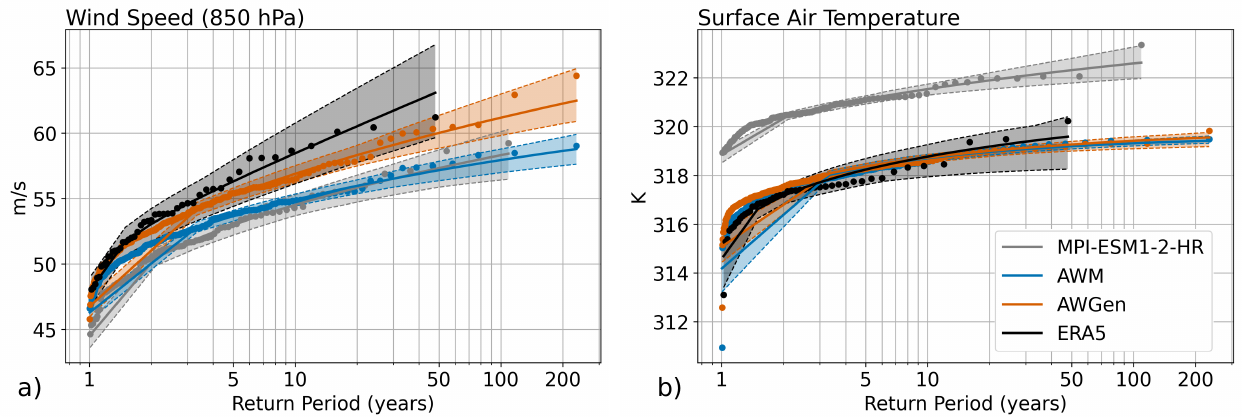}
    \caption{Return periods for surface air temperature and eastward wind at 850\,\unit{hPa} ArchesWeatherGen captures the return periods better than the other models when compared against the ERA5 reference.}
    \label{fig:extremes}
\end{figure}

In \ref{fig:extremes} a) we compare the wind speeds at 850\,\unit{hPa}. As observed in the histograms and the radial spectra, ArchesWeatherGen resolves localized oscillations better and matches the tails of the distributions more closely than ArchesWeather. The computed return periods match this observation. ArchesWeatherGen is closer to the return period curve computed from ERA5 while MPI-ESM1.2-HR and ArchesWeather simulate significantly weaker wind speeds at higher return periods.

In Fig. \ref{fig:extremes} b) we can observe that our models closely match the return periods of ERA5 for surface air temperature but also have a smaller slope. The discrepancy between our models and ERA5 might be a result of the weaker trend seen in the annual cycle in Fig. \ref{fig:tas_aw_mpi_era}. The confidence interval is narrower for our models (and for MPI-ESM1.2-HR) than for the ERA5 interval. The narrow confidence interval might be a consequence of the monthly mean forcings.

\subsection{+2\,\unit{K} and +4\,\unit{K} SST scenarios}
Following the AIMIP protocol, we uniformly add $+2 K$ and $+4 K$ to the sea surface temperature inputs over the ocean (with sea ice cover remaining unchanged) to evaluate the stability and response of the models to different scenarios of climate change in a simplified manner. This setup closely resembles the \textit{amip-p4k} experiment protocol from the CFMIP contribution to CMIP6 \citep{webb2017cloud}. 
However, the true response to these forcings is unknown, and a globally uniform increase in sea surface temperature, while keeping other variables the same, is physically unrealistic.

\begin{figure}[htbp!]
    \centering
    \includegraphics[width=0.9\textwidth]{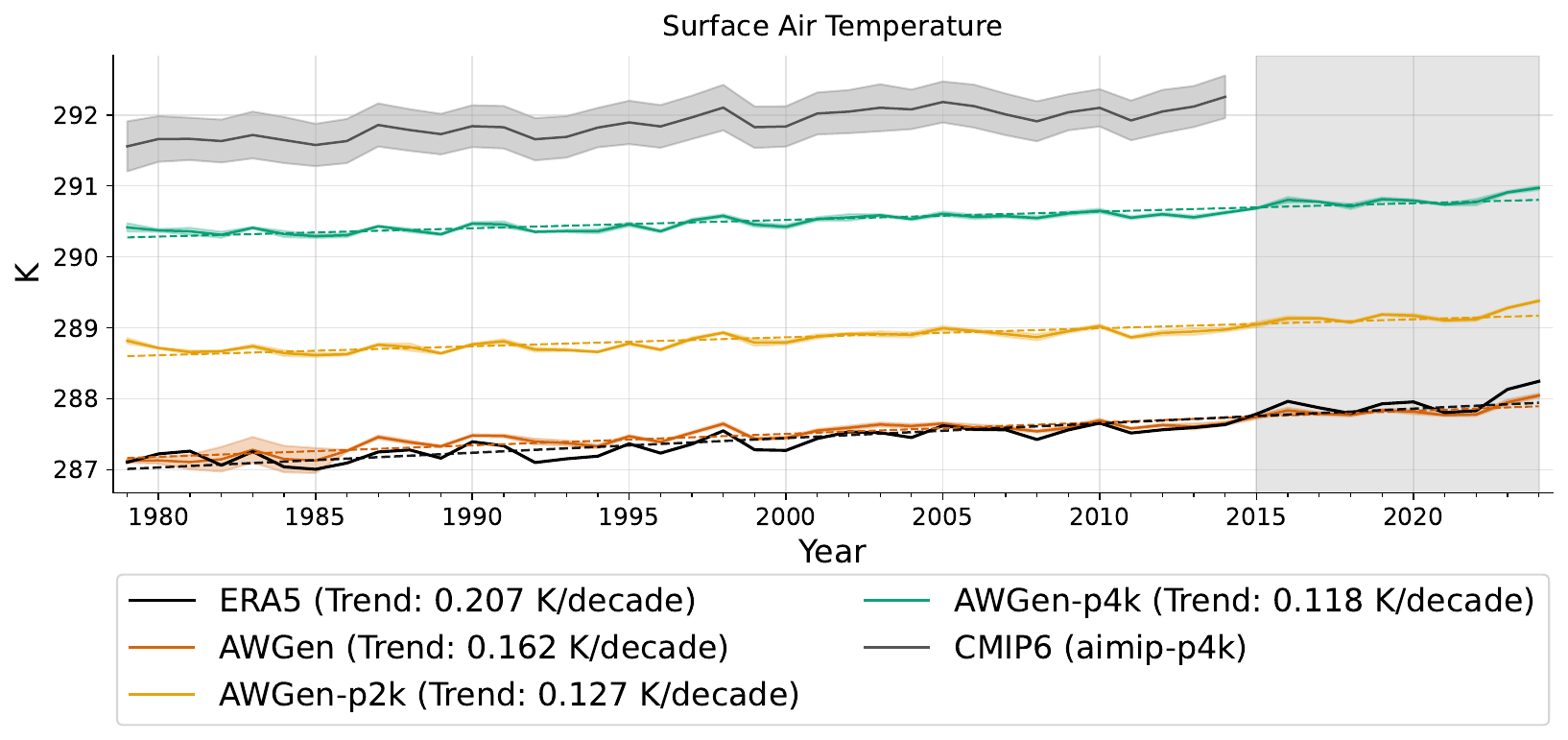}
    \caption{Global annual mean predictions for surface air temperature from ArchesWeatherGen 5-member ensemble compared against ERA5 ground truth (black) and CMIP6 amip-p4k ensemble (grey). When SST forcing is uniformly increased by +2 K (yellow) and +4 K (green), the model responds with shifted temperature trends.}
    \label{fig:p24K_trend}
\end{figure}

Figure \ref{fig:p24K_trend} shows that the ArchesWeatherGen ensemble remains stable despite out-of-distribution (OOD) forcings and responds with an increased surface air temperature with a proportional magnitude for the $+2 K$ and $+4 K$ forcing data. It is smaller than the $+2 K$ and $+4 K$ increase of ERA5's surface air temperature, respectively; although such a direct response is likely not realistic. It is also smaller than the average global response (4.48) from the CMIP6 ensemble of amip-p4k runs \citep{ringer2023global}, but we should keep in mind that the historical amip runs for these models are also biased compared to ERA5.

\begin{figure}[htpb!]
    \centering
    \includegraphics[width=1.0\textwidth]{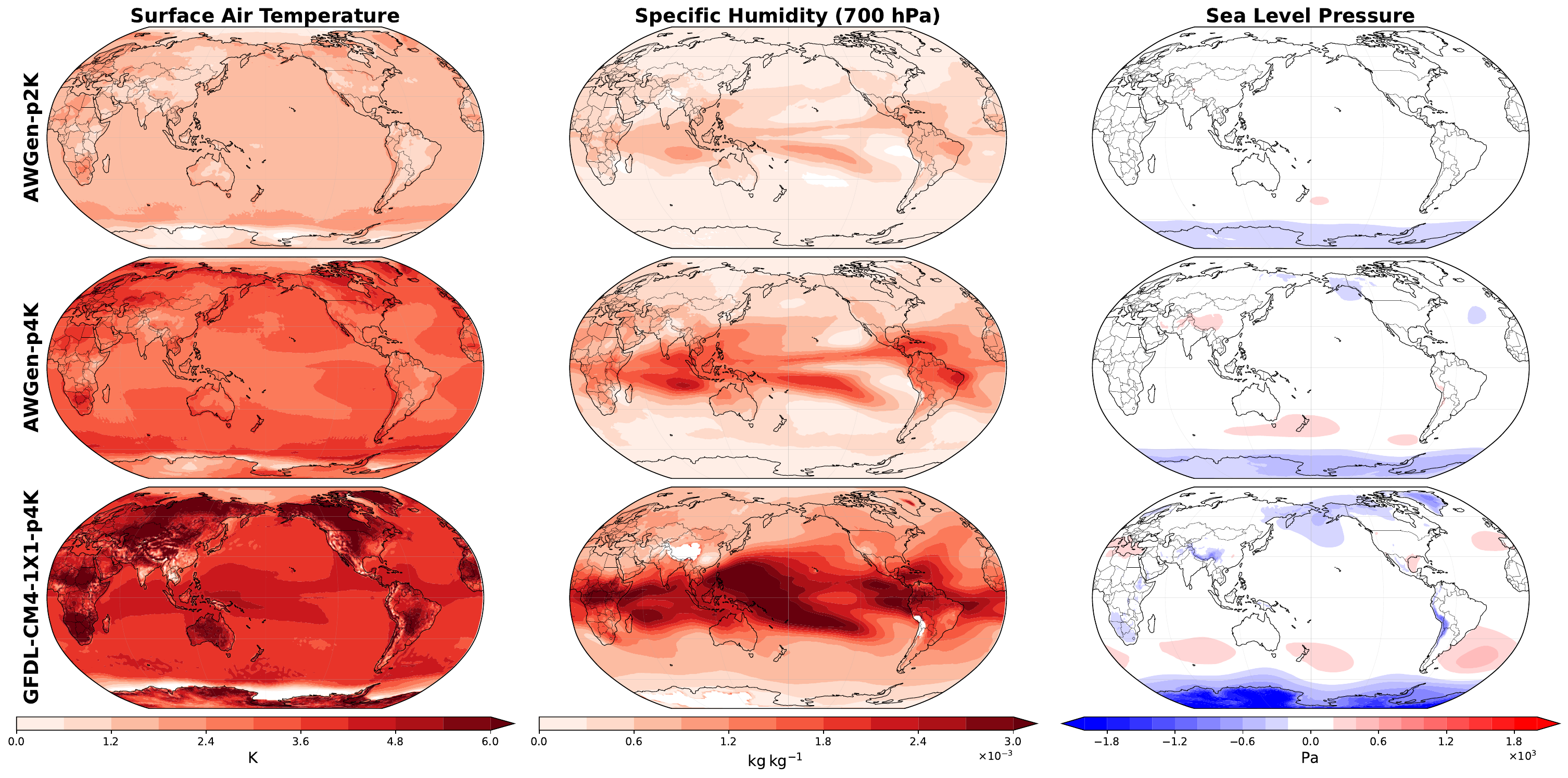}
    \caption{Spatial bias maps showing the global response of our model for surface air temperature,  specific humidity at 700\,\unit{hPa} and sea-level pressure. We also add GFDL as a reference. The model tends to produce consistently stronger warming in the southern hemisphere high-latitudes. Further, the model demonstrates that it is able to produce the higher humidity in the tropics expected for SST warming scenarios. For sea-level pressure, major difference can be observed of the Antarctic. }
    \label{fig:p4k_bias}
\end{figure}

\begin{figure}[htpb!]
    \centering
    \includegraphics[width=1.0\textwidth]{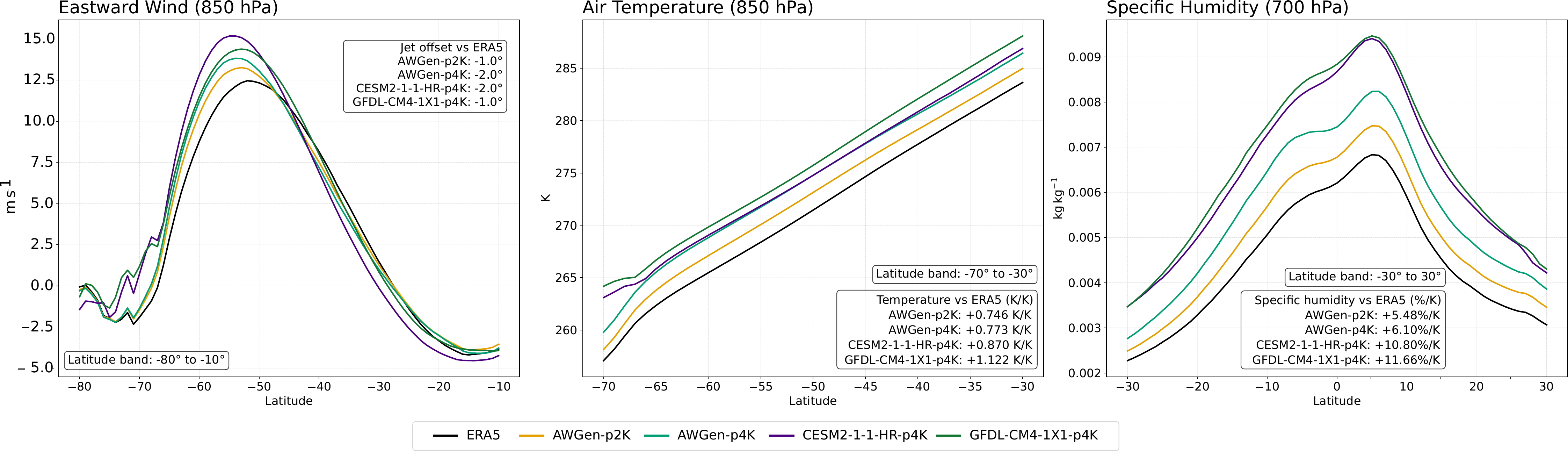}
    \caption{Latitudinal profiles for surface air temperature, eastward wind at 850\,\unit{hPa} and specific humidity at 700\,\unit{hPa}. For surface air temperature, we see that our models underestimate the trend, similar to the bias maps. For eastward wind, we see the expected northward shift of the jetstream and for specific humidity ArchesWeatherGen shows a response close to the Clausius-Clapeyron approximation.}
    \label{fig:p4k_latitudinal}
\end{figure}

We further assess the response to the heated SST forcings by plotting global bias maps (Fig. \ref{fig:p4k_bias}) and latitudinal profiles (Fig. \ref{fig:p4k_latitudinal}). Figure \ref{fig:p4k_bias} shows the global bias for ArchesWeatherGen, $+2K$, and $+4K$ for the GFDL model. For ArchesWeatherGen, an increased warming response can be found in the southern mid-latitudes over the entire pacific ocean for both the +2\,\unit{K} and +4\,\unit{K} scenarios. Stronger responses are also observed in the northern and southern parts of Africa and North America. The global response for GFDL is stronger over the entire globe compared to our models, with intensified warming over the landmasses. For specific humidity, the models simulate a significant increase around the equator, with particularly strong responses over the Pacific Ocean near the Niño 3.4 region, over the Indian Ocean, and over the Brazilian rainforest. Responses in sea-level pressure are most pronounced for all models starting from 60° South up to 90° South.

To further quantify the effects, we look in Fig.~\ref{fig:p4k_latitudinal} for the zonal-mean  surface air temperature, zonal-mean zonal winds at the level of 850\,\unit{hPa}, and zonal-mean specific humidity at 700\,\unit{hPa} compared to the ERA5 climatology. We show the two scenarios for ArchesWeatherGen and also include  the CESM2 and GFDL numerical models for the +4\,\unit{K} scenario from the CFMIP experiment. As noted above, numerical models cannot be taken as ground truth since the true physical response is unknown. As seen from the latitude-longitude maps, surface air temperature has a stronger increase at higher southern latitudes. The CFMIP models CESM2 and GFDL both show a response in which the increase in surface air temperature roughly matches the increase in sea surface temperature. ArchesWeatherGen slightly underestimates the increase, and more interestingly, the response following the +2\,\unit{K} case is close to the response of +4\,\unit{K}, indicating that capturing the out-of-distribution cases becomes harder as we depart further from the learned distribution. For eastward wind at 850\,\unit{hPa}, we observe that all models increase the magnitude of the southern jetstream and shift the stream poleward by at least 1\unit{\degree} compared to ERA5. This is the expected response. Finally, for specific humidity at 700\,\unit{hPa}, zonal-mean specific humidity is depicted around the equator where we observed the strongest responses from the spatial bias maps. We expect from the Clausius-Clapeyron relationship an increase of 7\,\unit{\%} per 1\,\unit{K} increase in temperature. ArchesWeatherGen yields a good approximation of it (+6.1\,\unit{\%}), while ArchesWeather underestimates the increase. Both physical climate models have a significantly higher increase.

We make some hypotheses about why our models' climate response to warmed SST forcings is smaller than that of the physical models. First, the deterministic ArchesWeather models predict the next state directly. Because of the regression to the mean effect, models conditioned on OOD forcings are biased to predict states closer to the training distribution's mean, which can explain why the climate response to OOD forcings is smaller than expected. We initially believed this effect would be dominant and that the model would not generalize to OOD forcings; yet, we still see a surprisingly good level of generalization. We see two possible ways to improve upon this. First, models could be trained to predict the difference between the next state and the current state, as is done, for instance, in GraphCast \citep{lam2023learning}. For such models, the bias would apply to the state increment rather than the state itself, and the effect would be alleviated. In earlier experiments, we found that the rollouts of such models are unstable, but there might be ways to stabilize them. We leave this for future research. Another possible option to improve the climate response is to increase the strength of the conditioning since we only used basic conditioning for training the models. A dedicated coupling module could be developed to ensure that the model predicts states that are compatible with its forcings.

\conclusions  

In this paper, we evaluated ArchesWeather and ArchesWeatherGen within the AIMIP protocol. While initially designed for weather forecasting, we adapted them as forced atmospheric models, notably by adding monthly-mean sea surface temperature and sea ice cover as input variables. 

We conducted an extensive evaluation of our models with respect to several climate patterns. The 45-years rollouts show that our models have long-term dynamical stability. They faithfully replicate historical climatology and annual variability patterns. In general, the climatology reproduced by our models closely matches the one of ERA5 as shown by the RMSE scores. While ArchesWeather achieves similar or even better results with respect to the RMSE, the CRPS shows differences for variables with high frequency content and extremes. There, ArchesWeatherGen significantly improves the results, especially for specific humidity. The radial spectrograms and histograms revealed the same results and further emphasized the advantage of the generative model. Our extended analysis of the reproduction of large-scale climate patterns revealed, that our models perform on par with physics based models. The return periods are only a proxy for analyzing extremes, but show statistics close to ERA5 and therefore suggest that our models are able to capture the tails over a long simulation period.

Simulating the observed climate is an important aspect of modern climate simulations but we further expect climate models to be robust to out-of-distribution forcings to forecast climate scenarios differing from the observed climate. Our models show strong response in specific humidity, close to the response expected in warming scenarios. Further, we observe a simulated poleward shift in the southern hemisphere jet stream which is also simulated by numerical climate models. Yet, the current response of our models to +2\,\unit{K} and +4\,\unit{K} SST warming, while significant, is smaller than expected. Since we only used basic conditioning to adapt weather models to forced atmospheric models, this could be improved with different design choices, like better conditioning methods or making a dedicated coupling module between the model's outputs and the forcing variables. Making these models fully generalizable to future climate change scenarios is an area of active research.

Our work emphasizes several limitations of these models. First, while the generative modeling paradigm provides a principled approach to generate an ensemble without requiring the perturbation of initial conditions or model parameters, the diffusion based approach is significantly more expensive. We leave the exploration of efficient generative methods for climate, like using diffusion distillation or CRPS optimization \citep{alet2025skillful} for future work. Second, modeling the ocean and the slower evolving processes in an unforced scenario is an open area of research. Therefore, coupling between the ocean and the atmosphere is likely necessary. Finally, including  other boundary conditions, for example, greenhouse gasses, is desired to enable the simulation of more realistic scenario cases.


\codedataavailability{Code to train and run models is available at \url{https://zenodo.org/records/20784771} \citep{couairon2026geoarches}. Model weights and configs will be made available on Hugging Face. ERA5 data for training is available for download on Copernicus (regridded to 1\unit{\degree} resolution with the internal Copernicus tool) and scripts to preprocess are made available in the repo above. The data needed to rollout models (initial conditions and forcings) are available at \url{https://zenodo.org/records/21952904}. The reference climate model data from AMIP used in this study is available from \url{https://wcrp-cmip.org/cmip-data-access/}. The specific amip-p4k runs used for the warming scenarios can be found via \url{https://doi.org/10.22033/ESGF/CMIP6.7536} 
\citep{https://doi.org/10.22033/ESGF/CMIP6.7536} and \url{https://doi.org/doi:10.22033/ESGF/CMIP6.8508}\citep{https://doi.org/10.22033/ESGF/CMIP6.8508}. We accessed all the data via the DKRZ compute cluster Levante. The models were regridded using the Climate Data Operators (CDO) \citep{schulzweida_2023_10020800}, \url{https://doi.org/10.5281/zenodo.10020800}. Forcing data is available at \url{https://zenodo.org/records/17065758} \citep{arcomano2025aimip}. The model outputs are available in the AIMIP Phase 1 data, publicly available via the DKRZ S3 endpoint (Download instructions available in \citep{henn_2026_21302180}. Code to convert model outputs into CMOR format as required by AIMIP and to evaluate it successively is available at \url{https://zenodo.org/records/21913147 } \citep{brunstein2026eval}. } 

\appendix
\section{Additional Experiments}    
\subsection{Response to forcings}\label{appendix:unforced}
In the paper, we add ocean surface forcings to simulate coupling with slow-moving processes in the ocean, which enables the model to replicate annual variability in the historical period. In order to understand their effect and verify that our model takes them into account, we compare our models to a configuration that is trained without these monthly mean forcings.

\begin{figure}[htpb!]
    \centering
    \includegraphics[width=1.0\textwidth]{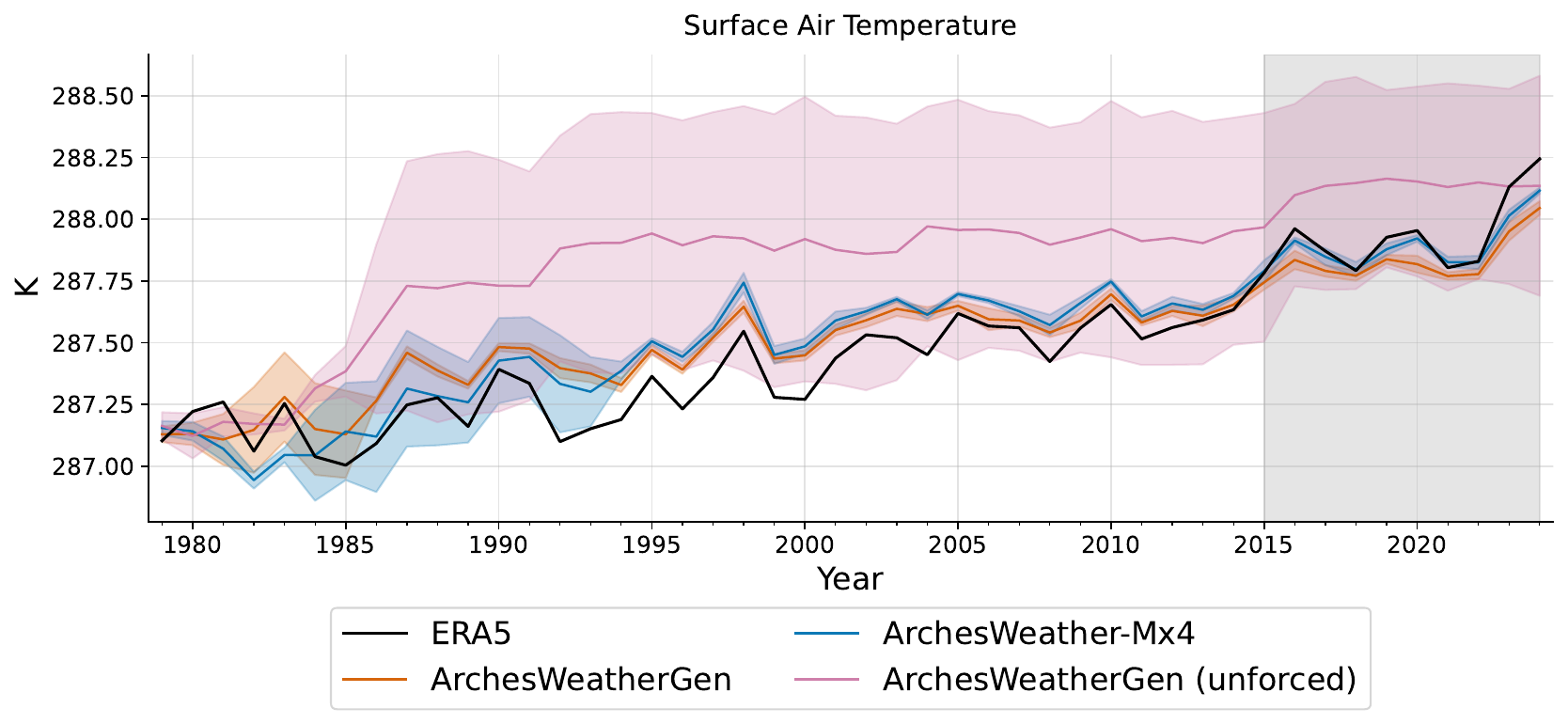}
    \caption{Global annual means of surface air temperature for ArchesWeatherGen, ArchesWeather, ArchesWeatherGenu, MPI-ESM1.2-HR and ERA5. The unforced model shows a stable cycle after a after its long spin-up, but with a strong bias. Furthermore, it does not pick up the increasing temperature trend like the forced models.}
    \label{fig:tas_aw_era_mpi_unforced}
\end{figure}

\textbf{Global annual mean trend} We look at the rollouts of the unforced model in Fig. \ref{fig:tas_aw_era_mpi_unforced}. Compared to the three forced models, the unforced configuration of ArchesWeatherGen requires a long spin-up period and stabilizes in a significantly biased climate relative to ERA5. Furthermore, while forced runs pick up the increased temperature trend, the unforced run does not, showing that, as expected, ocean surface forcings have a significant effect on the model's ability to replicate historical trends. 

\begin{figure}[htpb!]
    \centering
    \includegraphics[width=1.0\textwidth]{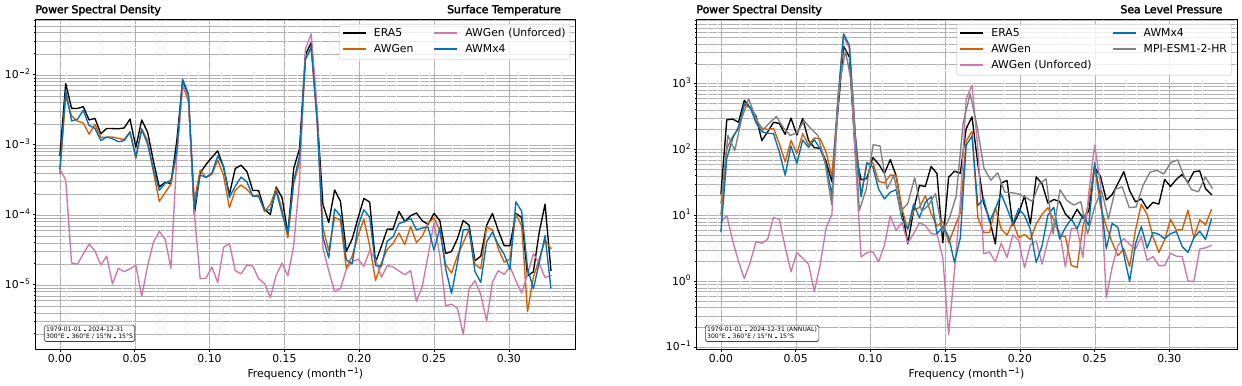}
    \caption{Power spectral densities of our models compared against ERA5. While the forced configurations capture even slow processes, the unforced configuration only closes the gap to ERA5 for higher frequencies.}
    \label{fig:psd_tos_psl}
\end{figure}

\textbf{Temporal Variability.} Figure \ref{fig:psd_tos_psl} shows the power spectral densities of ArchesWeather, ArchesWeatherGen, and ArchesWeatherGen U in an extended Nino3.4 region.  The depicted graphs showcase the influence of the forcing on the temporal resolution that can be represented. The unforced configuration has lower magnitudes for surface temperature and sea-level pressure and cannot capture the slower dynamics of these variables. The forced configurations, in contrast, have similar magnitudes to ERA5 and thus seem to infer the slowly evolving components from the monthly mean forcings. The results shown here also match the results presented in section \ref{section:enso}. Capturing such dynamics without forcings is a topic of ongoing research, as modeling these long-term dynamics is necessary to, e.g., predict warming trends or the El Niño Southern Oscillation.

\section{Ablations}
\subsection{Design choices}

We made several changes to the ArchesWeather backbone in order to run on 1\unit{\degree} resolution ERA5 data with added ocean surface forcings. Beyond the changes that we mention in Section \ref{section:architecture}, we detail several minor changes we made below:

\begin{itemize}
  \item \textbf{NaN interpolation}: Ocean surface variables (SST and SIC) only have valid values over the ocean. Therefore, we replace NaN values in the input with 0. During autoregressive rollouts, we also replace the same land values in the predicted ocean surface variables with 0 before passing the prediction as input.
  \item \textbf{Masked loss}: We apply a mask when computing the loss to ignore NaN values in the output ocean variables.
  \item \textbf{Time conditioning}: Instead of conditioning the model on the hour and month (of the current state $X_t$) as in the original paper, we replace it with the hour and day of the year.
  \item \textbf{Circular padding}: The decoder in ArchesWeather includes deconvolution layers to upsample back to the original resolution size. Instead of padding with 0, we apply circular padding to avoid having an arbitrary separation between the end and beginning of the image when the globe is spherical.
\end{itemize}

\subsection{Architecture ablations}\label{appendix:architecture}

\begin{figure}[htpb!]
    \centering
    \includegraphics[width=1.0\textwidth]{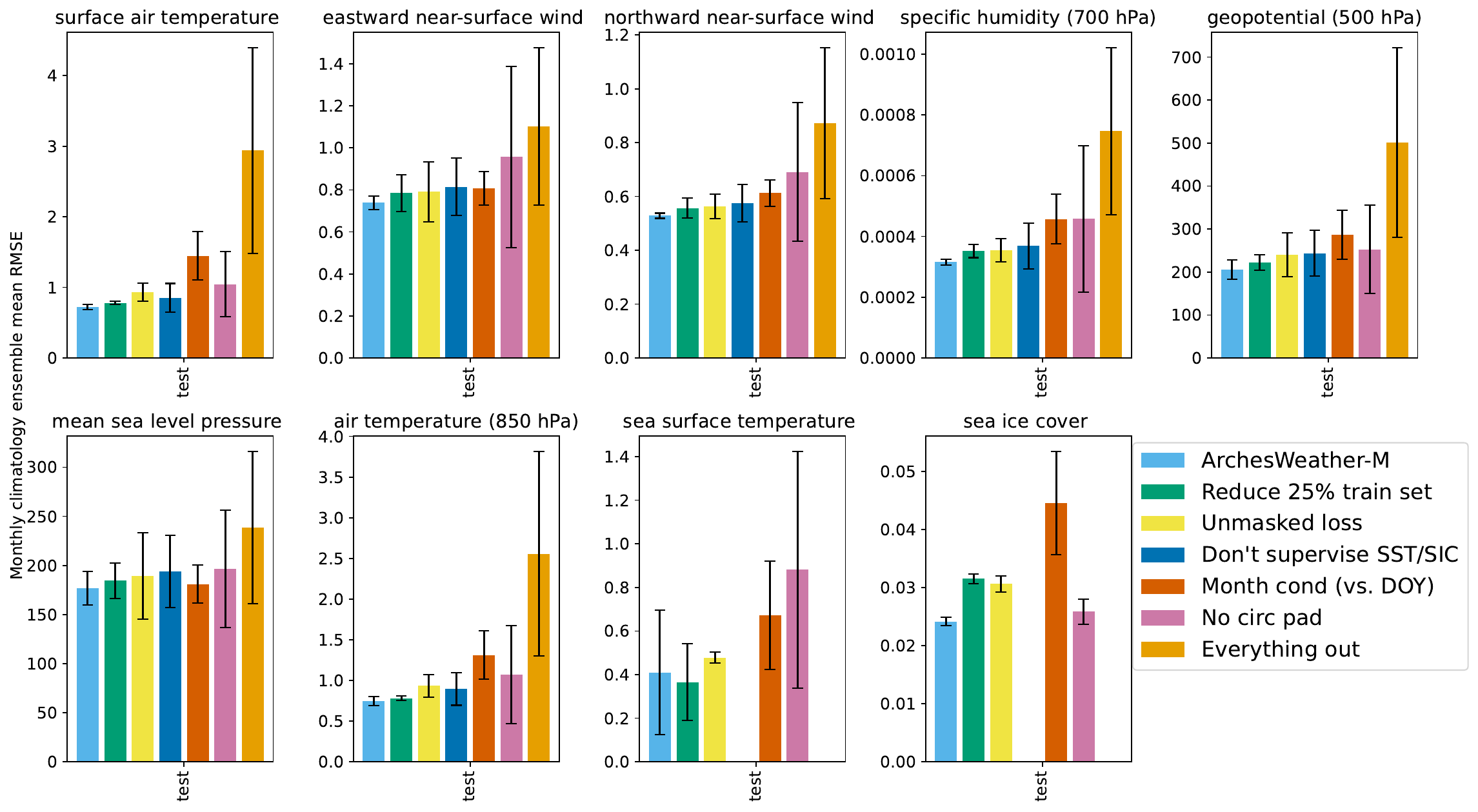}
    \caption{RMSE of monthly climatology for test period (2015--2024). The bar magnitude is the mean RMSE of 4 random seeds of ArchesWeather-M and the error bars are the standard deviation. The final configuration reproduces monthly climatology of ERA5 more faithfully than configurations without the final design choices.}
    \label{fig:rmse_ablations}
\end{figure}

We perform a set of ablations to isolate the improvement from each design choice. For each ablation configuration, we train 4 independent seeds of the single deterministic model, ArchesWeather-M. We observed that seed variability can be quite high (which may confound conclusions), but in general, we can conclude from Fig. \ref{fig:rmse_ablations} that the final configuration most faithfully reproduces mean climatology. We perform the ablations on the deterministic model as we found that, in general, biases carry over similarly to the generative model performance.

We perform each ablation by taking the final model and removing a single design choice. We run the following ablations:

\begin{itemize}
\item \textbf{Reduce 25\% train set}: The daily average dataset that we use could be considered redundant since we compute daily averages by applying a rolling window of length 4 and stride 1 to the 6-hourly ERA5 dataset. Hence, we experiment here with using a stride of 4 on the dataset (effectively using 1/4 of the training data.)
\item \textbf{Unmasked loss}: NaNs in the data are replaced in the case of target ocean variables, and the loss is computed over the full spatial field (sea and land values).
\item \textbf{Don't supervise SST/SIC}: We remove sea-surface temperature and sea ice cover from the output variables. We suspect that supervision on the daily averages of the ocean surface teaches the model to attend closely to monthly mean forcings.
\item \textbf{Month cond (vs. DOY)}: We use the time conditioning in the original ArchesWeather backbone, hour and month (instead of day of year).
\item \textbf{No circ pad}: We use zero-padding in the deconvolution layers in the original ArchesWeather backbone instead of circular padding.
\item \textbf{Everything out}: We apply all of the above ablations to one model.
\end{itemize}

\noappendix       




\appendixfigures  

\appendixtables   


\authorcontribution{RS, RB, GC, YH contributed to the code for adapting ArchesWeather to the AIMIP protocol (adding ocean forcings, downscaling to 1 degree etc.). RS and RB conducted the training, rollouts, and evaluation of the models. RS ran the ablations for validating the model. RB prepared training and validation data. AJ provided the CMOR code for the project and helped with evaluation. TR provided guidance on climate model evaluation. CL and CM provided guidance on the project.}

\competinginterests{No competing interests are present.} 


\begin{acknowledgements}
Thanks to Anastase Charantonis and Graham Clyne from the ARCHES team for discussions on climate modeling.
Robert Brunstein is funded by the Deutsche Forschungsgemeinschaft (DFG, German Research Foundation) – 537063406.
\end{acknowledgements}


\bibliographystyle{copernicus}
\bibliography{references}

\end{document}